\documentclass[preprint,12pt]{elsarticle}

\usepackage{amssymb}
\usepackage{amsthm}
\usepackage{graphicx}        
\usepackage{multicol}        
\usepackage[bottom]{footmisc}
\usepackage[english]{babel}
\usepackage{float}

\usepackage{amsfonts}
\usepackage{amsgen}
\usepackage{amsbsy}
\usepackage{amsmath}

\usepackage[colorlinks=true,linkcolor=blue,citecolor=red,urlcolor=cyan]{hyperref}
\usepackage{stmaryrd}
\usepackage{subfigure}
\usepackage{epsfig}
\usepackage{color}
\usepackage{lscape}
\usepackage{mathenv}
\usepackage{hyperref}
\usepackage{fancyhdr}
\usepackage{textcomp}
\usepackage{listings}
\usepackage{gensymb} 

\usepackage[normalem]{ulem}

\journal{Physical Review Fluids (PRF)}

\begin{document}

\begin{frontmatter}

\author{Onofrio~Semeraro\fnref{label1}}
\ead{semeraro [at] ladhyx.polytechnique.fr}
\author{Fran\c cois~Lusseyran\fnref{label2}}
\author{Luc~Pastur\fnref{label2}}
\author{Peter~Jordan\fnref{label3}}

\address[label1]{LadHyX, CNRS, \'Ecole polytechnique -- Palaiseau, FR}
\address[label2]{LIMSI -- CNRS, Orsay, France}
\address[label3]{Institut PPRIME, CNRS, Universit\'e de Poitiers, ENSMA -- Poitiers, FR}

\title{Qualitative dynamics of wavepackets in turbulent jets}

\begin{abstract}
It has long been established that turbulent jets comprise large-scale coherent structures, now more commonly referred to as ``wavepackets'' \citep{jordan2013wave}. These structures exhibit a remarkable spatio-temporal organisation, despite turbulence.

In this work we analyse, from a qualitative point of view, the temporal dynamics of axisymmetric wavepackets educed, experimentally, from subsonic iso-thermal jets. We use the data presented by \cite{breakey2013near}, where time-series of the wavepackets are extracted at different streamwise locations. A thorough analysis is performed; statistical tools are used for estimating the embedding and correlation dimensions characterising the dynamical system. System identification is used for computing nonlinear surrogate models. Finally, control-oriented linear models are computed.

The goal of the contribution is to assess the extent to which non-linear models are necessary, or appropriate, for description of the temporal wave-packet dynamics and to provide a complementary perspective to the current modelling.
\end{abstract}

\end{frontmatter}

\section{Introduction}\label{sec:intro}
Following the early studies of Moll\"o-Christensen \citep{mollo1960sound,mollo1964experiments,mollo1967jet}, a considerable body of work has been devoted to exploring the nature of organised motions that {are} observed in turbulent jets \citep{Crow-Champagne,lau1972intrinsic,Michalke-JFM-1975,armstrong1977coherent,moore-JFM-1977}. It was hypothesised early on that this component of the flow might be understood in terms of an instability of the turbulent mean \citep{Crow-Champagne,michalke1971instability,bechert1975amplification,cohen1987evolution,crighton1976stability,plaschko1979helical}; and the importance of such flow structures for sound radiation {has been} suggested {in} numerous studies \citep{michalke1970wave,michalke1971expansion,bishop1971noise,Crow-PhysSoc-1972,Crighton-Theory-1975,Michalke-JFM-1975,crighton1990shear}.

\begin{figure}
\includegraphics[width=\textwidth,clip=]{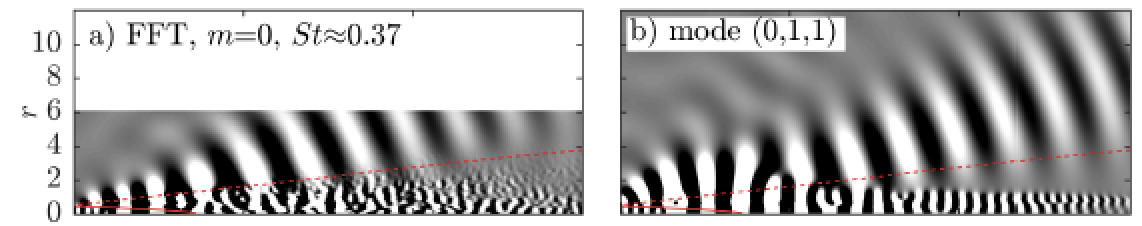}
\caption{Left: realisation taken from Large Eddy Simulation (axisymmetric component of pressure at $St=0.37$). Right: Global mode $(m,St)=(0,0.37)$. Courtesy of \cite{schmidt2016super}}\label{fig:wavepacket}
\end{figure}

It is only more recently, however, thanks largely to progress in theory, numerical simulation and experimental diagnostics, that it has been possible to explore these hypotheses in a comprehensive manner. Exhaustive comparison of the results of theory with experimental measurements has confirmed that the average characteristics of coherent structures in turbulent jets are remarkably well described by solutions of the Navier-Stokes equations {linearised about the turbulent mean} \citep{suzuki2006instability,cavalieri2013wavepackets,jordan2014modeling}; {these solutions are synonymous with globally stable modal solutions \citep{nichols2011global,garnaud2013preferred,schmidt2016super}, which, physically, amount to hydrodynamic waves that are convectively amplified in the upstream region of the flow, but become neutrally stable and then decay farther downstream.} It is for this reason that they have been given the denomination `wavepacket' \citep{jordan2013wave}. {An example of a modal solution, from \cite{schmidt2016super}, is shown in figure \ref{fig:wavepacket}, where it is compared with a realisation taken directly from the Large Eddy Simulation \citep{bres2015large} that provided the mean flow. The wavepackets can be seen to comprise a hydrodynamic component, with an amplitude envelope as described above, and an acoustic component that takes the form of a directive beam radiating to shallow angles.}

{Considering the Reynolds number of the jet, $Re=1\times10^6$, and the fact that it issues from a nozzle with fully turbulent boundary layers, the agreement between the LES realisation and the linear global mode is striking. One is compelled to ask how, despite the non-linear, orderless character of the turbulence that dominates the fluctuation energy of this flow, such organisation can exist and how a linear model can capture so many of its features.}

{However, something less clear from the images 
is the considerably lower acoustic efficiency of the linear wavepacket. As shown in a number of recent studies \citep{zhang2014just,jordan2014modeling,baqui2015coherence}, the intensity of the soundfield of the linear solution, once the hydrodynamic fluctuation levels of LES and model have been matched, can be as much as 40dB below that of the LES. Such a large discrepancy, in the face of such compelling organisational similarity, is intriguing and the subject of a number of ongoing research efforts, which we summarise briefly in what follows.}

{The present understanding is that the} linear dynamics do not contain the acoustically important degrees of freedom: the average wavepacket does not generate the average sound \citep{cavalieri2011jittering,cavalieri2011using,kerherve2012educing,jordan2014modeling,zhang2014just}. The essential sound-producing motions are those associated with higher-order statistics of the wavepacket dynamics. These motions have been denoted `jitter' by \citep{cavalieri2011jittering}, and would appear to be underpinned by non-linearity \citep{jordan2014modeling,zhang2014just,towne2015stochastic,tissot2016sensitivity}.

\medskip
The nature of this non-linearity remains to be clarified, and it motivates the present study. Do wavepackets jitter on account of non-linear wave-wave interactions \citep{sandham2006nonlinear,suponitsky2010linear}, i.e. is the non-linearity of an intrinsic kind?  Or is it rather an extrinsic, stochastic forcing of linear waves by background turbulence that leads to the activation of these higher-order degrees of freedom  \citep{towne2015stochastic,tissot2016sensitivity}? Or might both extrinsic and intrinsic mechanisms be at work? {In this work, in light of the foregoing considerations and questions, we are interested in low-dimensional projections of the complete wavepacket dynamics.}

\smallskip
{A further issue, that motivates both the modelling efforts evoked above and the work we undertake in this paper, is that of control. Given the success of linear models in predicting both the average wavepacket structure, in frequency space, and the real-time evolution in an experimental context \cite{sasaki2015real}, is it legitimate to ask if the tools provided by linear control theory might be sufficient? Such possibilities are presently being explored by \cite{sasaki2016closed}. But if it were to be necessary to directly manipulate the more subtle dynamics associated with the energetically-unimportant-but-acoustically-essential degrees-of-freedom, then a non-linear control framework would be required (machine learning etc. \cite{gautier2015closed}). In both cases the question of the dimension of the space spanned by the dynamics must again be asked: is this small enough for control to be realistic in an experimental context?}

\smallskip
{And, finally, there is the issue of the physical interpretation of POD modes, which some recent results show may correspond to the highest gain structures of the linear Navier-Stokes operator subject to stochastic external forcing by turbulence. With this in mind, the various analyses of the dynamics will be performed using both raw and POD-filtered data.}

\bigskip
Consideration of {the above issues} via qualitative dynamical analysis has, to the best of our knowledge, not previously been attempted. We therefore consider the same turbulent jets studied by \citep{cavalieri2012axisymmetric,cavalieri2013wavepackets,breakey2013near}, using a variety of non-linear signal-processing and system identification techniques, in order to explore  dynamics of low-frequency wavepackets. Particular attention is given to pressure signals dominated by fluctuations at Strouhal numbers $St<0.2$, where linear models are found to perform most poorly.

Our analysis starts with a minimal description of the experimental setup in Sec.~\ref{sec:expdata}; we refer to \cite{breakey2013near} for a detailed description. The remainder of the paper is organised in two parts. The behaviour of the time-series obtained from the experimental campaign is discussed in Sec.~\ref{sec:tseries}, where statistical tools and spectral analysis are introduced for a preliminary characterisation  of the temporal dynamics. 

In the second part, we pursue an alternative approach by analysing reduced-order models of the time-series, based on system identification; the approach is introduced in Sec.~\ref{sec:sysID}, while the dynamical analysis is reported in Sec.~\ref{sec:rom1} and Sec.~\ref{sec:rom2} . More details on the techniques are given in \ref{sec:appA}. The manuscript finalizes with a discussion of the results and concluding remarks in Sec.~\ref{sec:conclu1}.

\section{Experimental data}\label{sec:expdata}
The study presented in this paper is based on the analysis of the temporal behaviour of axisymmetric wavepackets educed from iso-thermal turbulent jets. In particular, we consider the near-field pressure fluctuations measured by \cite{breakey2013near}. For a detailed description of the measurement setup, data acquisition and post-processing the reader can refer to that paper. We here limit ourselves to a brief presentation of the data, in order to highlight those features, which are relevant to the analysis we perform.

\begin{figure}
\centering 
{\includegraphics[width=0.495\textwidth]{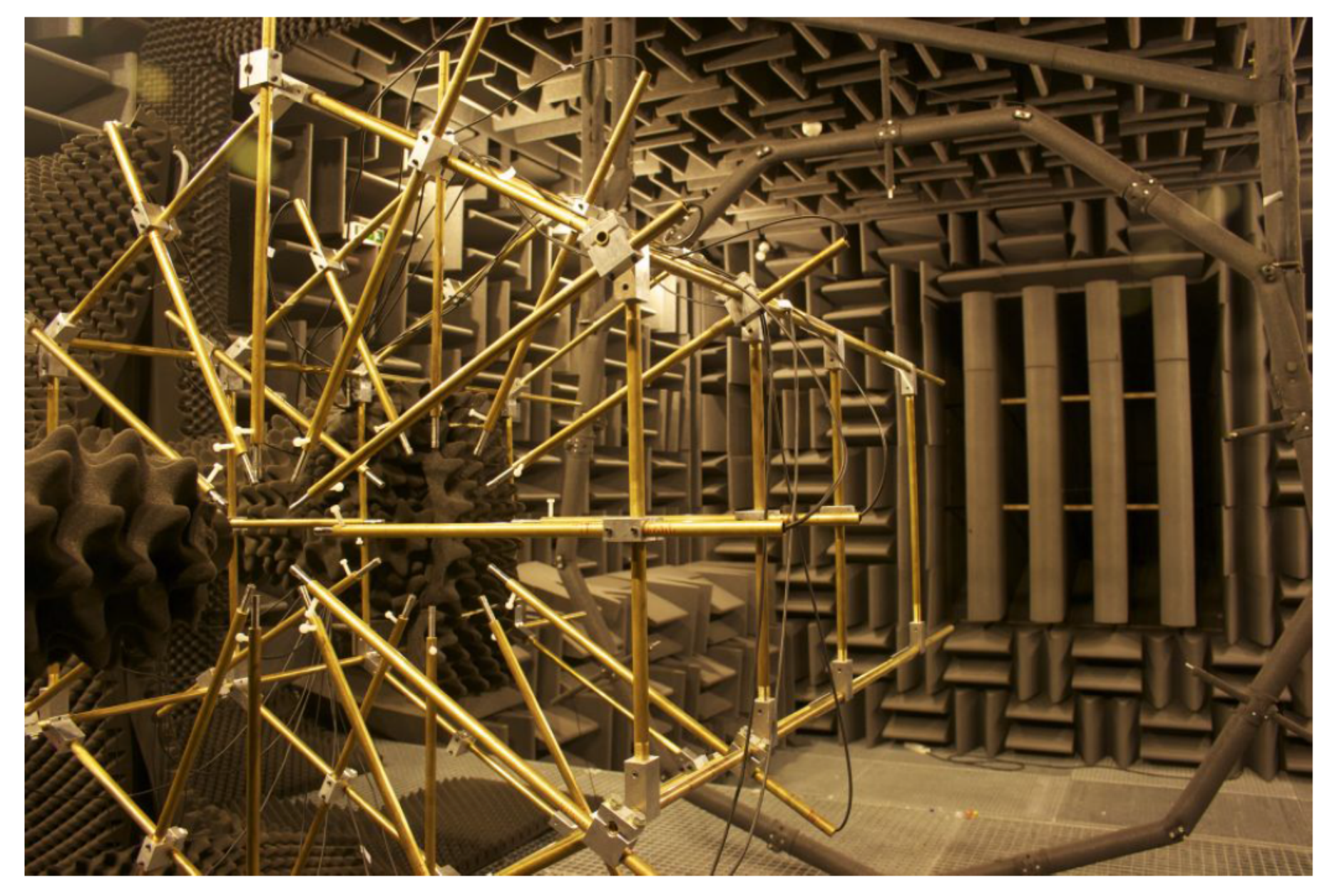}}
{\includegraphics[width=0.495\textwidth]{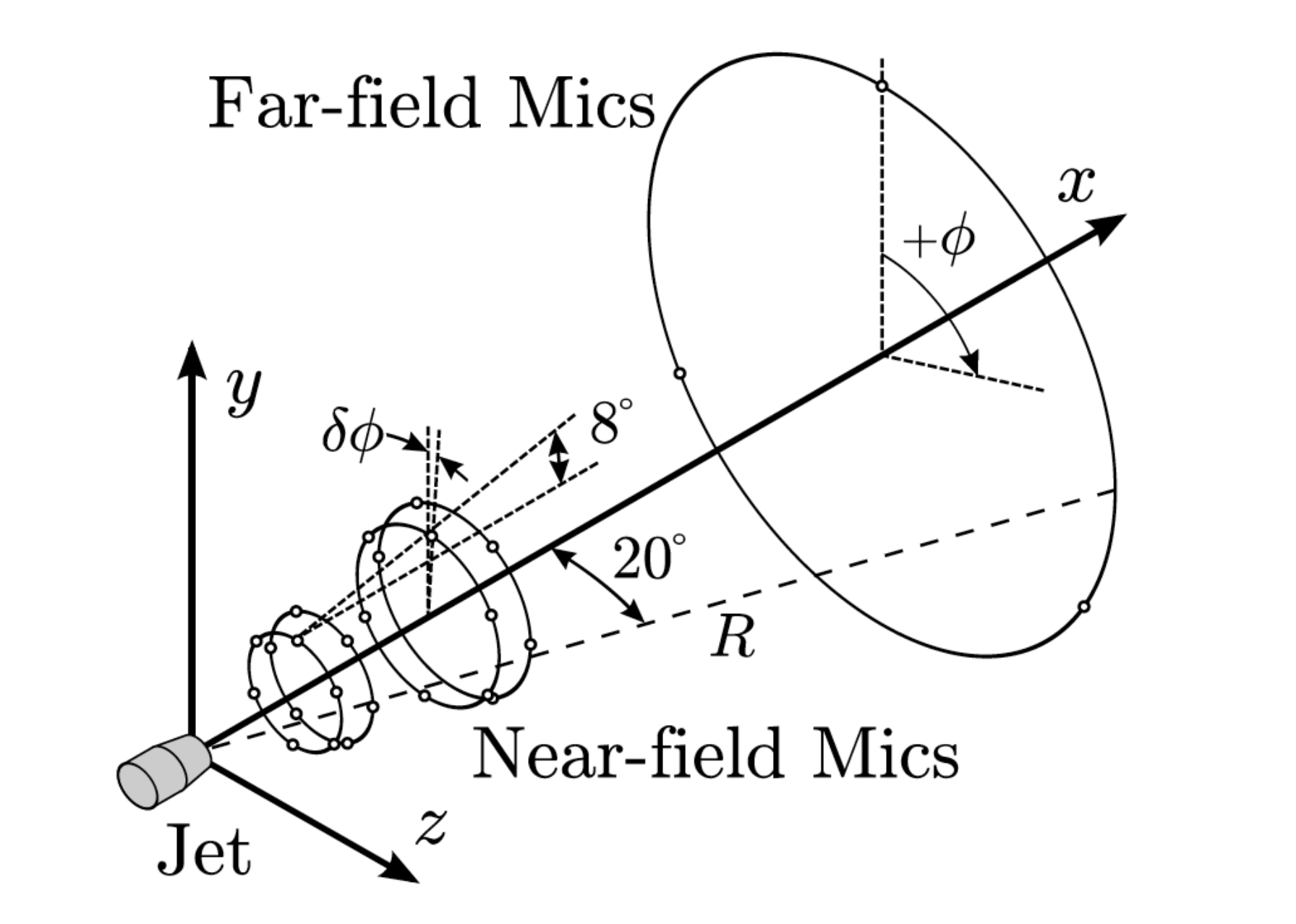}}
\caption{Overview of the experiment, adapted from \cite{breakey2013near}.}
\label{fig:setup}
\end{figure} 
\subsection{Experimental setup}
We consider an unforced, isothermal, subsonic jet issuing from a round nozzle with a fully turbulent boundary layers. The flow is thus very different to the transitional jet studied by \cite{narayanan2013dynamical}. The setup is shown in Fig.~\ref{fig:setup}$a$, and a schematic of the near-field microphone array used for eduction of the axisymmetric wavepacket signature is sketched in Fig.~\ref{fig:setup}$b$.  The experiments were carried out in the anecho\"ic free jet facility at the \emph{Centre d'Etudes A\'{e}rodynamiques et Thermiques (CEAT)}, Institut Pprime, Poitiers (FR). The jet Mach number ranged from $Ma=0.4$ to $Ma=0.6$.  In what follows, the diameter of the nozzle,  $D=0.05m$, is taken as the reference length-scale, while the exit velocity, $U$, is the reference speed. Time is scaled as $\tau=tU/D$. The corresponding Reynolds number $Re={\rho U D}/{\mu}$ ranges between $4.2\times 10^5 < Re < 5.7\times 10^5$, where $\rho$ is the density and $\mu$ the air viscosity. The potential core length of the jet ranges between $5<x/D<5/5$.

\begin{figure}
\includegraphics[width=0.99\textwidth]{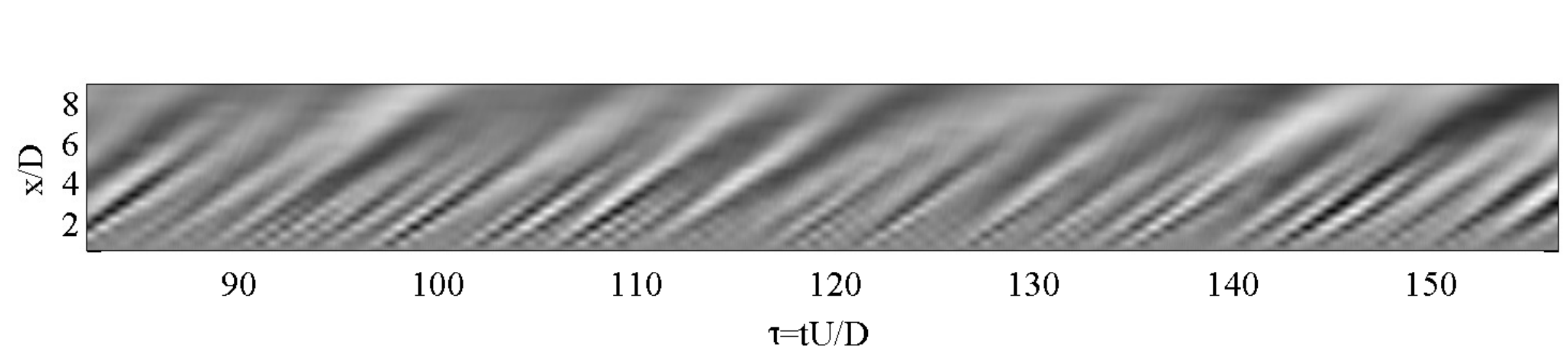}
\caption{The near-field signal is shown as a function of the distance from the nozzle exit ($x/D$) and the time $\tau$, scaled with the velocity $U$ and the diameter $D$. The measured quantity is the pressure related to the axisymmetric mode $m=0$. The time-series are obtained from the POD reconstruction described in \cite{breakey2013near}.}
\label{fig:spatiotemp}
\end{figure} 
Azimuthal rings of six microphones recorded pressure fluctuations from the near-field of the jet, on a conical surface. An azimuthal ring of three microphones recorded the far-field fluctuations. In this work, we focus on the near-field fluctuations; two different datasets are analysed, obtained from two measurement campaigns.

\begin{enumerate}
\item The first set of data is obtained by simultaneous measurements at $7$ different streamwise locations in the near-field. At each location, an azimuthal ring, each with $6$ microphones, is positioned. The array thus comprises a total of $42$ microphones, placed on a conical surface in the near-field of the jet, as shown in Fig.~\ref{fig:setup}$b$. The spacing between the rings is $x=0.75D$. At each $x/D$ location a Fourier-series decomposition is performed in the azimuthal direction.
\item  The measurements of the second campaign are performed using a rig of $4$ equi-spaced, azimuthal rings, which are displaced in the streamwise direction over the range $0.5D<x<8.9D$ with a streamwise resolution of $x=0.4D$. This set of measurements allows a more detailed computation of the two-point flow statistics, providing the cross-spectral-density matrix, an eigen-decomposition of which provides POD modes. 
\end{enumerate}
This POD basis is used for the projection of the data from the first campaign. In this way, a larger domain along $x/D$ is spanned. The temporal dynamics of the axisymmetric mode $m=0$ is shown in Fig.~\ref{fig:spatiotemp}, as a function of time $\tau$ and the streamwise direction. This space-time picture is complementary to the frequency-space realisation shown in Fig.~\ref{fig:wavepacket}. Both show clearly the highly organised nature of wavepackets in these high-Reynolds-number, fully turbulent jets.

\medskip
Our study focuses on the temporal dynamics of the axisymmetric mode $m=0$. The analysis is carried out by considering the \emph{time-series} extracted at given streamwise locations, for each of the datasets. By definition, a time-series is a sequence of equi-spaced data points, function of time. For our case, the sampling rate is $10^5 Hz$, and the total number of points for each series is $N=2.4\times 10^5$.

In the following section, preliminary analysis of the temporal behaviour is performed by means of spectral analysis and estimation -- using non-linear tools -- of the embedding and correlation dimensions associated with these wavepackets.

\section{Part I: Time-series analysis}\label{sec:tseries}

In this section, we consider both the original data and the time-series post-processed by {means} of proper-orthogonal decomposition (POD). In particular, we study the runs at $Ma=0.6$ and Reynolds number $Re=5.7\times 10^5$, at different locations along the streamwise direction ($1.25<x/D<5.75$, $0.5<x/D<8.9$).

\medskip
The main goal in this section is to characterise the time--series from a dynamical-system point of view. Although we are considering a fully turbulent case, we study a projection of the system on a set of linear, axisymmetric wavepackets. Thus, the working hypothesis is that the resulting temporal dynamics is low-dimensional. In principle, a dynamical system can be qualitatively analysed by reconstructing the -- unknown -- underlying attractor from a sequence of observables of its state, for instance the time-series. This idea is at the heart of the Takens' theorem \cite{takens1981detecting}, a delay embedding theorem providing the conditions under which the reconstructed dynamics preserves the properties of the original system. In particular, the theorem ensures a diffeomorphism between the original -- unknown -- phase space and the reconstructed phase space. In that sense, {there are two} crucial parameters: i) the minimal embedding dimension $m$, \emph{i.e} the dimension of the {projection phase space} of the reconstructed {dynamics}; ii) the correlation dimension $d_2$, a measure of the fractal dimension of an attractor (see \cite{grassberger1983characterization}). According to these values {one} can evaluate the dimensions of the attractor of the system and -- if possible -- study its projection.

\medskip
In general, the available time series are univariate; thus, the embedding is {typically} the first step for the qualitative analysis of a dynamical system. Different procedures of embedding are available. For instance, a {singular-value} decomposition of the Hankel matrix based on univariate time-series can be performed, where the dimensions of the rectangular matrix are the length of the time series $N$ and the maximum embedding. A second alternative consists of computing higher-order derivatives of the time-series, such that each of the {derivatives corresponds} to one of the {columns} of the embedding vector. In this work, we rely on the most classical approach: once a maximal embedding dimension $m$ is chosen, each coordinate is obtained by a time shift $\Delta t$; thus, introducing the time series $y(t)$, the embedding vector is obtained as
\begin{eqnarray}
\mathbf{Y}\left(m,\Delta t\right) = \left[y(T), y(T+\Delta t), y(T+2\Delta t), ..., y(T +m\Delta t) \right],
\end{eqnarray}
where $T$ indicates the time-span; note that the last channel will have a total shift of $m\Delta t$ with respect to the first one. Once an embedding vector is obtained, we need to {identify} the minimal embedding dimension $m$ that allows {determination of} the dimension of the (hypothesised) attractor. A small $m$ indicates the possibility of unfolding the attractor. A popular method for estimating the minimum embedding dimension is the \emph{false nearest algorithm}, \cite{kantz2004schreiber, cao1997practical}. 

%
%
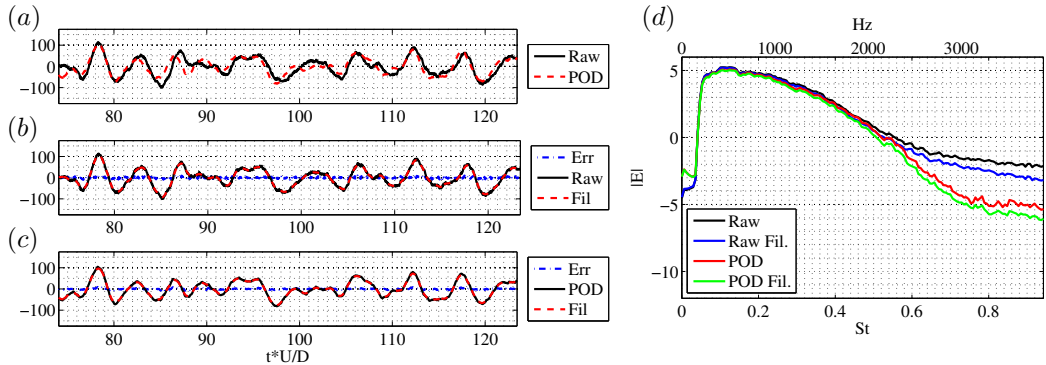
\begin{figure}
\centering 
\includegraphics[width=0.999\textwidth]{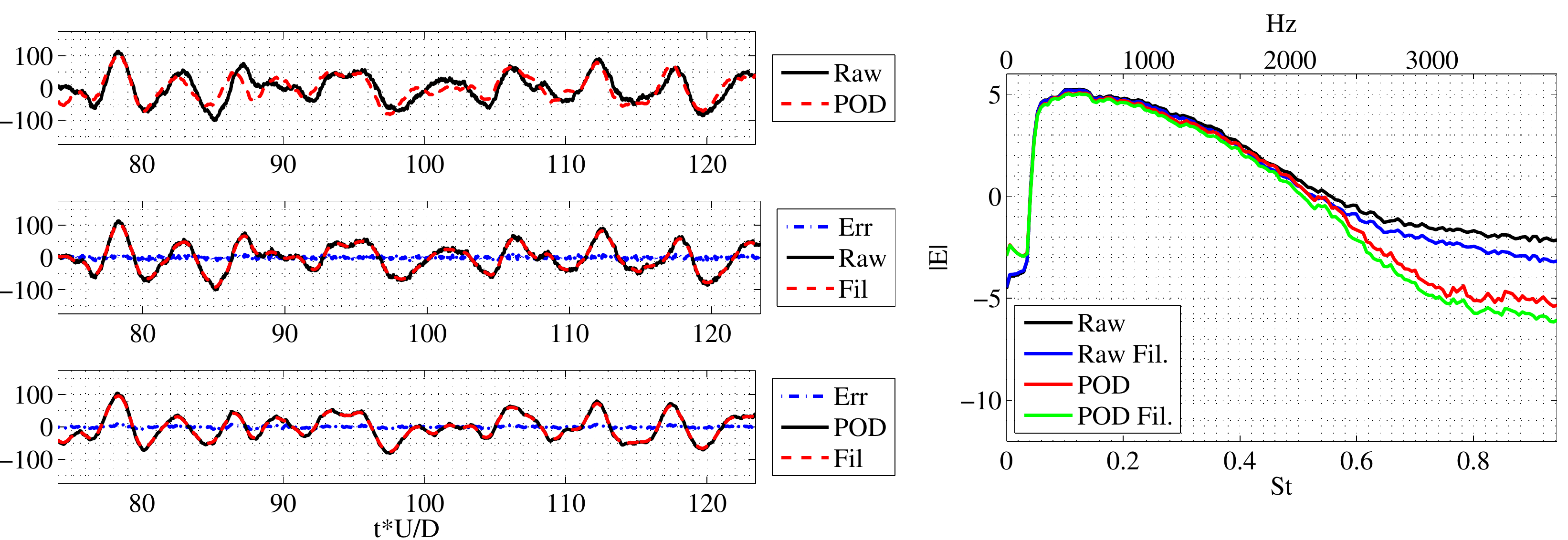}
\begin{picture}(0,0)(0,0)
    \put(-200,142.5){\footnotesize{($a$)}}
    \put(-200,100.5){\footnotesize{($b$)}}
    \put(-200, 58.5){\footnotesize{($c$)}}
    \put( 40 ,142.5){\footnotesize{($d$)}}
\end{picture}
\caption{Temporal behaviour for the mode $m=0$ at $x/D=5.7$. In the inset $(a)$ the original signal (\emph{TS-Raw}) is compared to the result of the POD post-processing (\emph{TS-POD}), at a location $x/D=5.75$. A second non-linear filter is applied to both the series as shown in the insets $(b)$ and $(c)$, where the original time-series (black solid line) is compared to the filtered one (red dashed line); the resulting error is shown as a blue dashed-dotted line. Finally, in the inset $(d)$ the spectrum of the four time-series is shown. A difference between \emph{TS-Raw} and the \emph{TS-POD} is observed for $St>0.6$, while good agreement is obtained at $St$ numbers relevant for our analysis.}
\label{fig:tseries1}
\end{figure} 

\medskip
The second dimension of interest is the correlation dimension $d_2$, providing an {estimate of} the fractal dimension of an attractor. This quantity is related to the embedding dimension $m$ by the Takens' theorem, stating that $m\le 2d +1$  (\cite{takens1981detecting}).

\medskip
In what follows, we adopt statistical tools described from the theoretical point of view in the book by Kantz \& Schreiber \cite{kantz2004schreiber}. A thorough discussion of these techniques is beyond the scope of this paper: we will focus mainly on the validation of the results. Unfortunately, these statistical methods are capable of producing results also in the presence of purely random time-series. Thus, the validation has a twofold goal: assessing the robustness of the results, and confirming the determinism of the dataset.

%
%
\subsection{Filtering and spectral analysis}\label{sec:filtering}
%
%
\begin{figure}
\centering 
\includegraphics[width=0.45\textwidth]{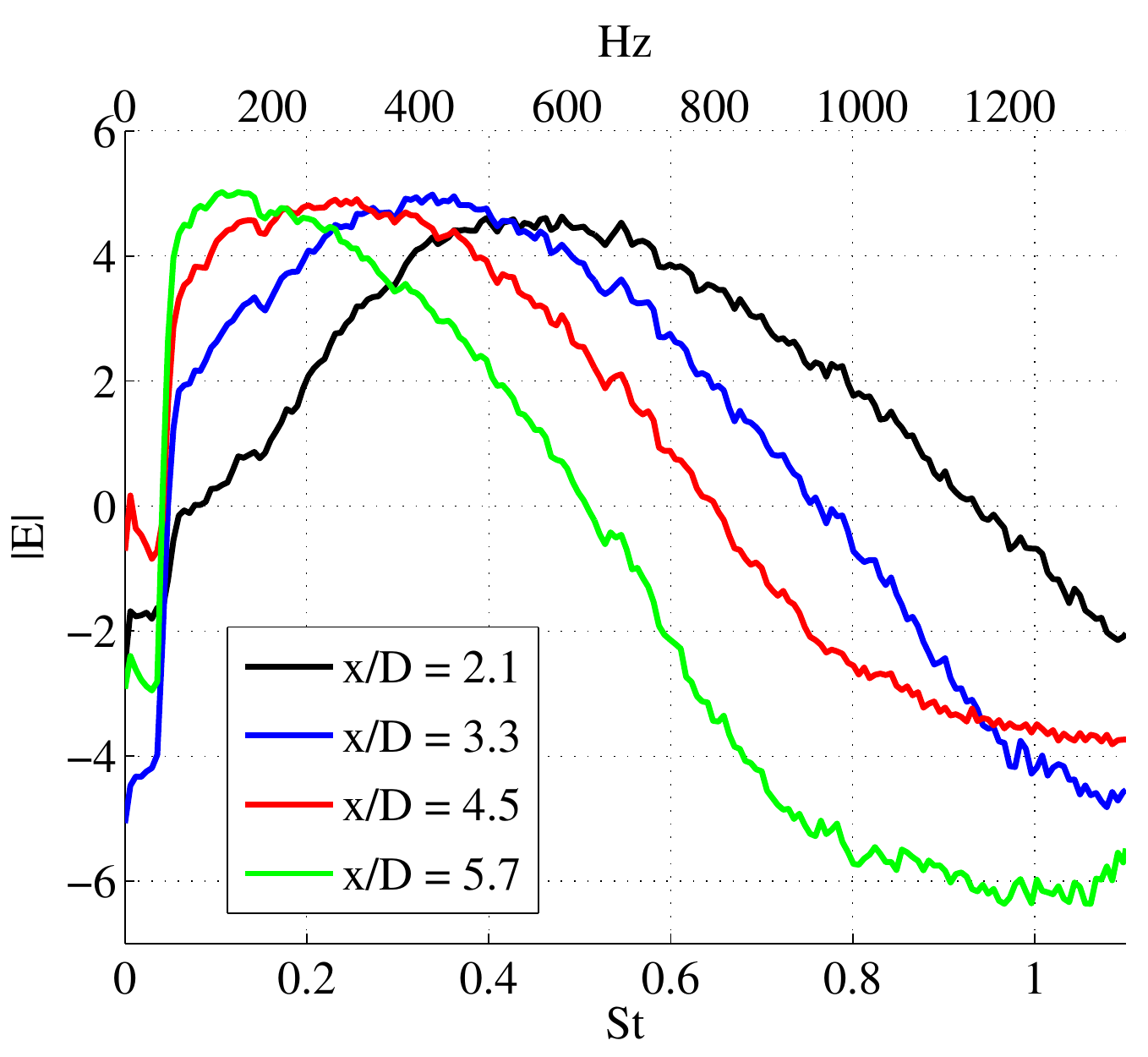}
\caption{Spectral analysis for the series based on POD projection, at different locations along $x/D$, namely $x/D=2.1$, $x/D=3.3$, $x/D=4.5$ and $x/D= 5.7$. It can be observed that the peak shifts from $St\approx0.50$ to $St\approx0.15$ when considering downstream locations.}
\label{fig:pwelchseries}
\end{figure} 
In order to analyse the time-series, we first {perform} non-linear noise reduction. Indeed, noise is a dominant, limiting factor for the embedding procedure \cite{kantz2004schreiber}. The de-noising algorithm is a moving-average filter, applied along the trajectory identified in the embedding space, on the assumption that the dynamics is {continuous}. In Fig.~\ref{fig:tseries1}, the time-series is shown at $x/D=5.7$. In particular, we are interested in assessing to what extent the applied filters modify the essential dynamics of the time-series. In the following, for {ease of} discussion, the original time-series {is} denoted \emph{TS-Raw} while data filtered using {POD} as {the projection} basis {is} denoted \emph{TS-POD}. 

\medskip
In Fig.~\ref{fig:tseries1}$a$, the \emph{TS-Raw} data are compared with the \emph{TS-POD} set in {the} time-domain. A quantitative assessment is reported in Fig.~\ref{fig:tseries1}$d$ using Welch's power spectral density (PSD) estimate. At the significant Strouhal numbers, $St=fD/U$, {it can be see} that the POD filter does not {strongly} modify the dynamics of the time-series. Differences can be observed only at $St>0.6$. At lower $St$, we observe the cut-off due to {high-pass filtering of the original data applied to remove energy below the anecho\"ic cut-off frequency of the windtunnel}. The two {data} sets are filtered using the non-linear filters in Fig.~\ref{fig:tseries1}$b$ and Fig.~\ref{fig:tseries1}$c$. By definition, the non-linear filter does not act on specific bandwidths, but on the whole spectrum, while preserving the foliation of the attractor in the embedded space. Welch analysis confirms this behaviour: in Fig.~\ref{fig:tseries1}$d$ we can observe that the non-linear filter acts on the whole range of $St$ numbers. The standard deviation of the removed noise is $\sigma = 0.075$ and $\sigma = 0.063$, for \emph{TS-Raw} and \emph{TS-POD}, respectively.

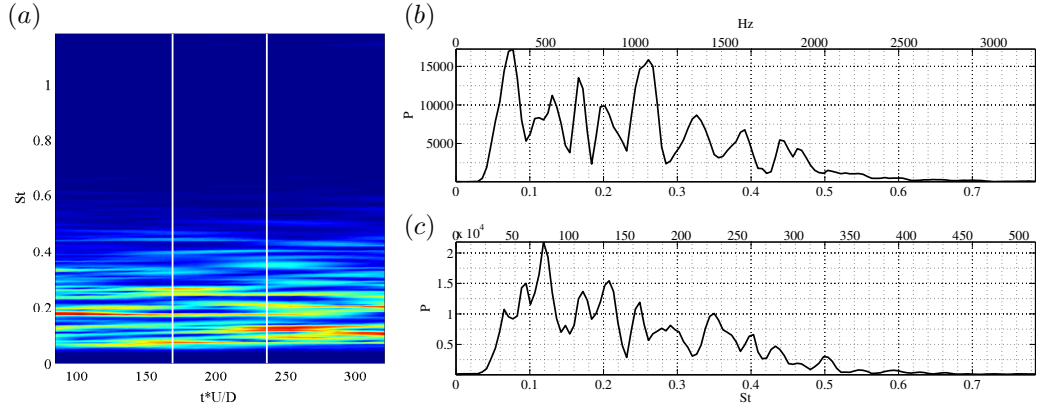
\begin{figure}
\centering 
\includegraphics[width=0.37\textwidth]{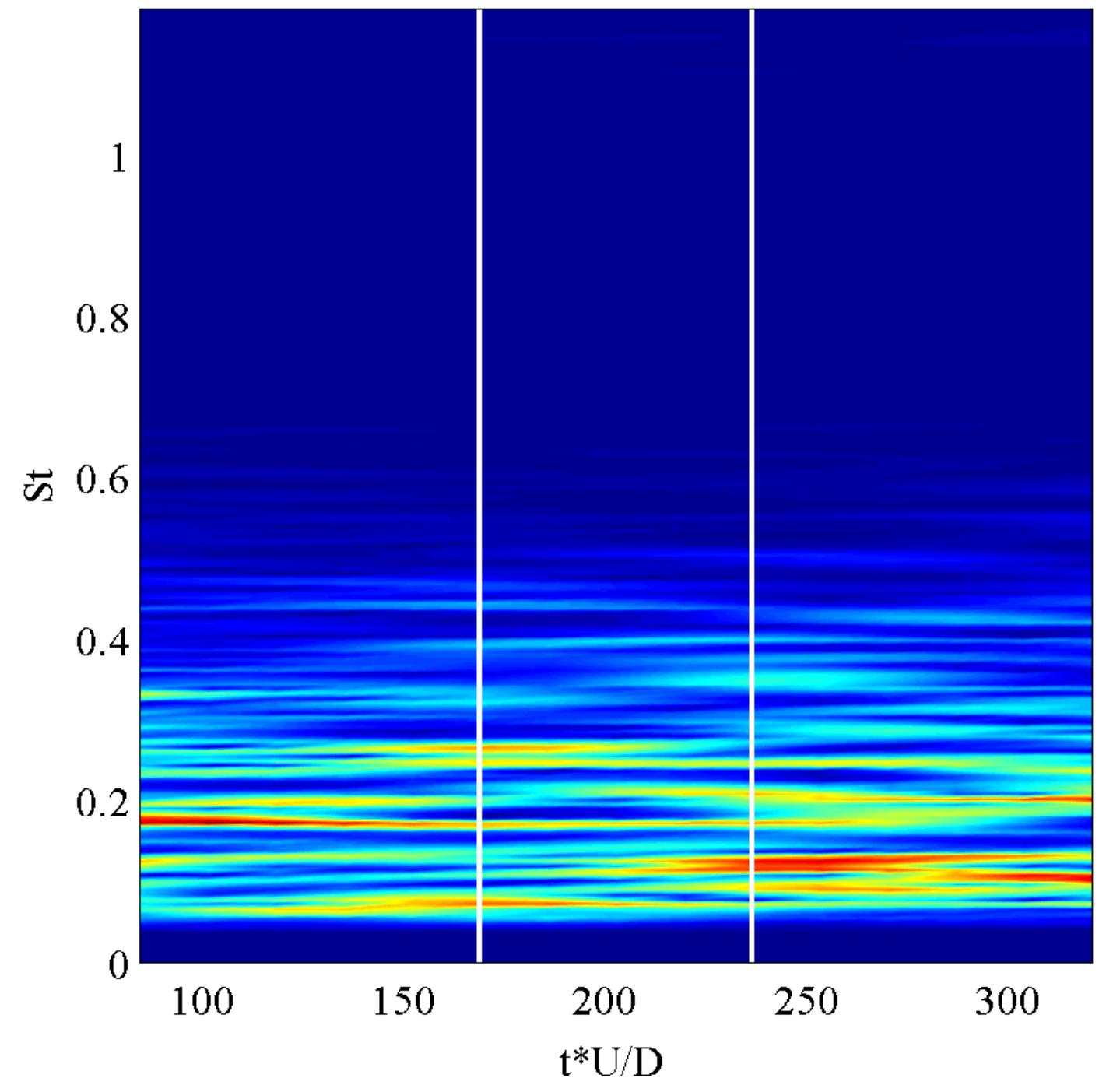}
\includegraphics[width=0.62\textwidth]{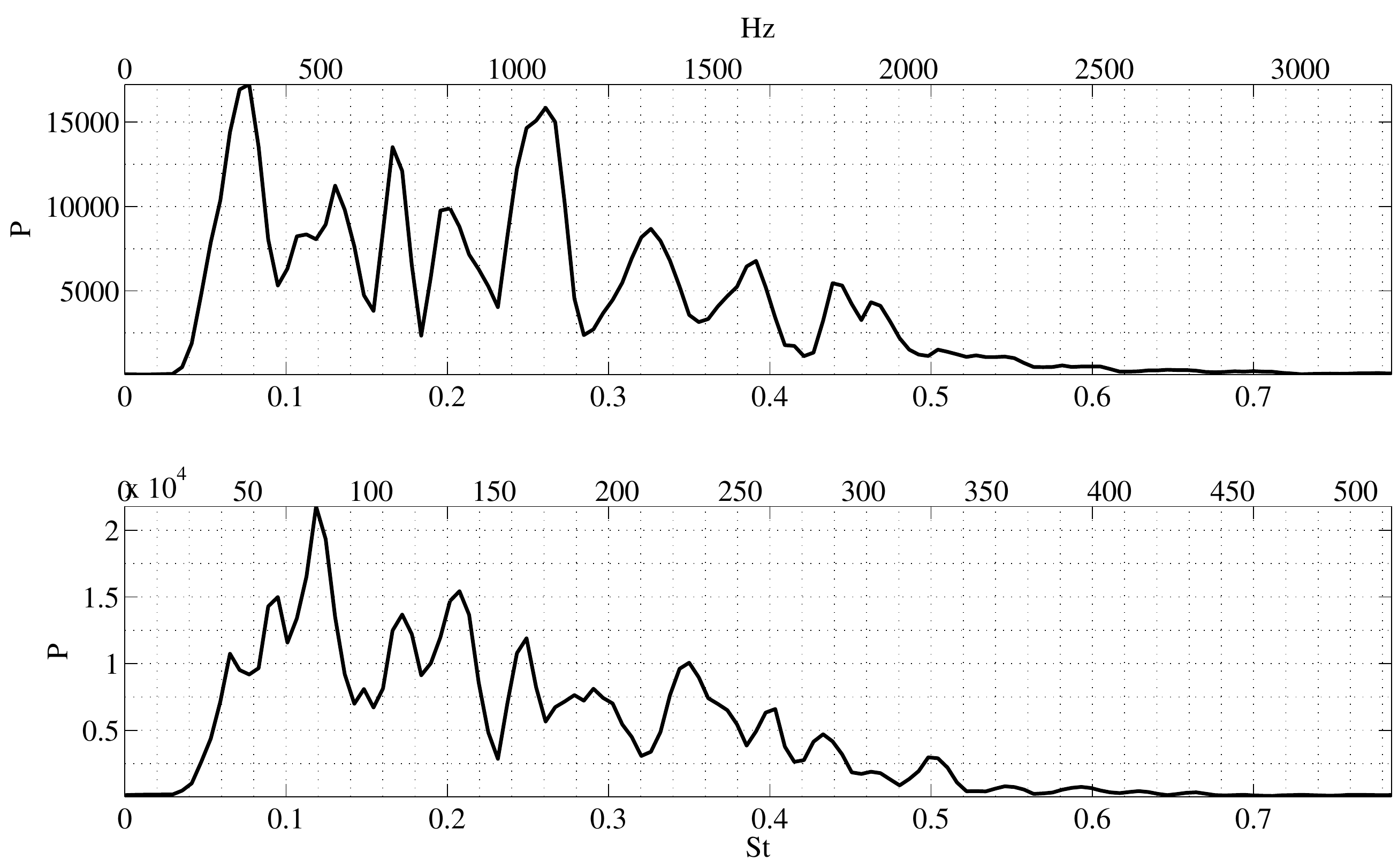}
\begin{picture}(0,0)(0,0)
    \put(-195,160){\footnotesize{($a$)}}
    \put(-45,160){\footnotesize{($b$)}}
    \put(-45, 80){\footnotesize{($c$)}}
\end{picture}
\caption{Spectrogram for the time-series based on the POD projection at $x=5.7$. In the insets $(b)$ and $(c)$, the power spectrum is shown as a function of the Strouhal number $St$ at $\tau\approx 170$ and $\tau\approx 240$.}
\label{fig:spectrogram}
\end{figure} 
%
%
%

The spectral content of the \emph{TS-POD} data at $x/D=[2.1,\,3.3,\,4.5,\,5.7]$ is compared in Fig.~\ref{fig:pwelchseries}. We observe that the maximum fluctuation levels move progressively to lower frequencies as the observation position moves downstream, from $St\approx 0.50$ at $x/D=2.1$ to $St\approx 0.15$ at $x/D=5.7$. A change in the maximum amplitude can also be observed. An alternative analysis is provided by the spectrogram, shown in Fig.~\ref{fig:spectrogram}$a$ for the \emph{TS-POD} at $x/D=5.7$. The graph is in two dimensions: the horizontal axis represents the time scale, while the $St$ number is reported along the vertical axis. The amplitudes are shown as a colour map: at each instant, the intensity of the fluctuation is shown as a function of $St$. The Welch estimate can be regarded as an average along the time-span of the results shown in the spectrogram. In particular, in Fig.~\ref{fig:spectrogram}$b-c$, two instants are shown at $\tau\approx 170$ and $\tau\approx 240$, respectively. The instantaneous amplitudes of certain frequencies in the bandwidth can be observed; this behaviour is a time-frequency view of the organised behaviour manifest in space-time in Fig.~\ref{fig:spatiotemp}.
These behaviours are the signature of the amplifier nature of the flow: wavepackets are hydrodynamic instability waves that exist on the turbulent mean flow. These waves acquire initial amplitudes and phase at upstream stations -- either from disturbances issuing from the turbulent motions within the nozzle, or from distributed turbulence that acts as a volume force -- and evolve in the downstream direction according to the stability properties of the linear operator, possibly subject also to distributed forcing from background turbulence; this last issue is one that is presently being studied.

%
%
\subsection{Embedding dimension and correlation dimension}
Assessment of the correlation dimension, $d_2$, and embedding dimension, $m$, is performed for both \emph{TS-Raw} and \emph{TS-POD} datasets, filtered non-linearly. The results are collected in Fig.~\ref{fig:dimension}, as a function of the streamwise direction $x/D$.

\subsubsection{Minimal embedding dimension}
\begin{figure}
\centering 
\includegraphics[width=0.6\textwidth]{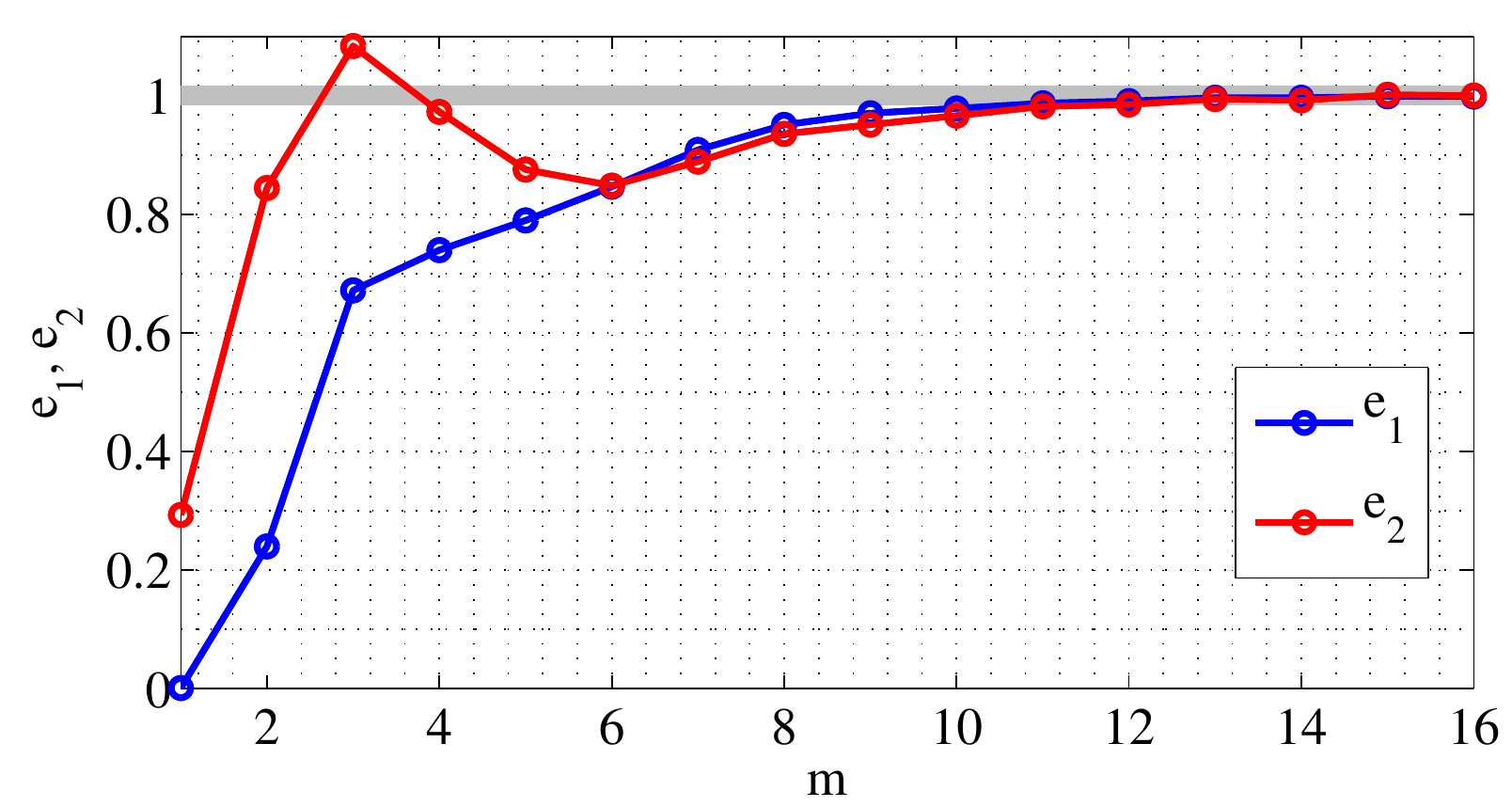}
\caption{Example of minimum embedding dimension $m$ estimation. The POD-filtered data are used, in $x/D=5.7$. The saturation of the parameter $e_1$ (blue curve) indicates the minimal embedding $m$; the complementary parameter $e_2$ (red curve) confirms the determinism of the dataset, as $e_2\neq 1$ for $m<13$.}
\label{fig:e1e2}
\end{figure}

The minimal embedding dimension $m$ is estimated by using the algorithm described by Cao in \cite{cao1997practical}. The algorithm does not strongly depend on the length of the time series and it is capable of estimating the embedding dimension also for time series describing high-dimensional attractors. More importantly, it allows to clearly determine whether the signal is deterministic or stochastic, as shown in Fig.~\ref{fig:e1e2}. The dimension $m$ is estimated by analysing the behaviour of the parameter $e_1$: the saturation indicates the minimal embedding dimension. We chose it using a threshold value $\delta=0.99$, giving $m=13$. The quantity $e_2$ explicitly accounts for the correlation: random series will be characterised by $e_2(m)=1$, for any $m$. For deterministic data, $e_2$ cannot be a constant. In our case, we find that the values of $e_2$ are not constantly unitary, as shown in Fig.~\ref{fig:e1e2}: this is a first clue that we are analysing deterministic time series.

%
%
\medskip
A critical aspect is the choice of the embedding delay $\Delta t$. In principle, the embedding dimension $m$ is independent of the time delay; in practice, the minimum embedding dimension estimate may depend on this choice and needs to be verified case by case. In Fig.~\ref{fig:dimension}, we identify a confidence interval for the $m$ estimates; the upper bound corresponds to a time-delay $\Delta t=0.21$, while the lower bound is obtained for $\Delta t=0.29$. The choice of the embedding delay is done by analysing the auto-correlations. $N=1.0 \times 10^4$ points have been used for the computations, although we verified that $m$ is scarcely affected by this choice.

The confidence intervals are shown using shadowed areas. Each point corresponds to the value where the parameter $e_1$ saturates, as discussed in Fig.~\ref{fig:e1e2}. Dotted curves identify the minimal embedding dimension computed by averaging the results obtained with $N=[1.0,\, 1.5, \, 2.0]\times 10^4$ points and $\Delta t=0.21$. It can be observed that the \emph{TS-Raw} series is characterised by a rather constant value of $m$, oscillating between a minimum value $m=12$ at $x/D=5$ and maximum values $m=13-14$ further upstream, at $x/D=2$. This estimate follows closely the upper bound of our confidence interval. The lowest value is found at $x/D=5.7$, for the runs at $\Delta t=0.29$.

The behaviour of the \emph{TS-POD} series is coherent with the original data, although we found higher values. This apparently counter-intuitive behaviour is discussed later (see Sec.~\ref{sec:discuemb}). For values $x/D>5.7$, the minimal embedding dimension is practically constant until $x/D=8$; however, note that in this region the data are extrapolated, so relevant dynamics carried by higher order POD modes is missing.

\subsubsection{Correlation dimension}
%
%
\begin{figure}[h]
\centering 
\includegraphics[width=1\textwidth]{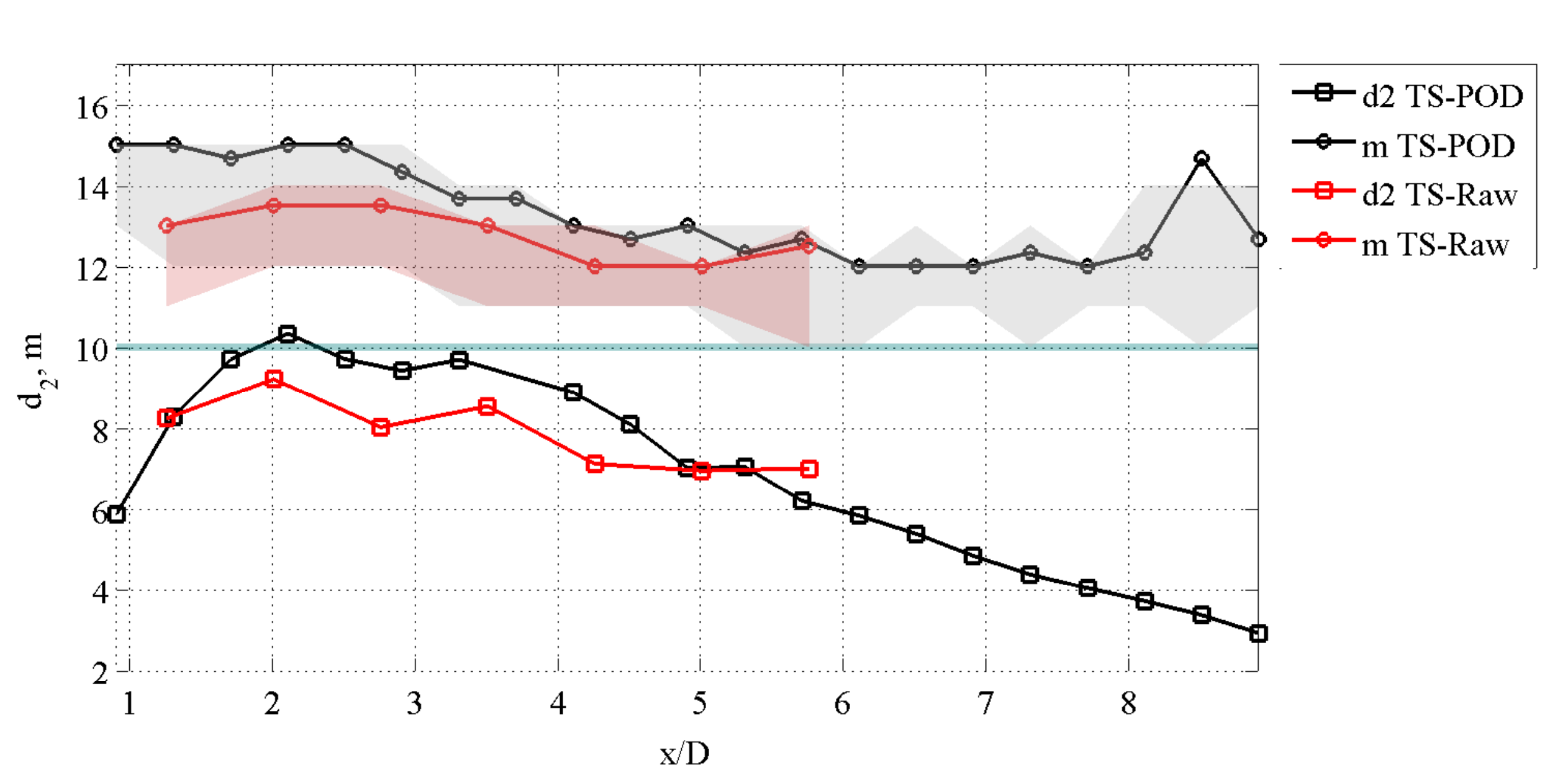}
\caption{The estimates of the embedding dimension $m$ and the correlation dimension $d_2$ are summarised as function of $x/D$ for both the datasets under investigation. The minimal embedding dimension $m$ is computed for two different time-delays, defining the confidence intervals indicated with shadowed areas: a smaller time of embedding, $\Delta t=0.21$, defines the upper limit, while a longer time, $\Delta t=0.29$, is related to the lower bound. The red area is obtained for the \emph{TS-Raw} series, while the grey one is associated with the results of the dataset \emph{TS-POD}. The nominal estimate for $m$ (dotted curves) is obtained by averaging the results of different runs. The correlation dimension $d_2$ is indicated with a squared-dotted black curve and a squared-dotted red curve for \emph{TS-POD} and \emph{TS-Raw}, respectively. A green line indicates the Eckmann-Ruelle limit: values of $d_2$ above this bound are not physical.}
\label{fig:dimension}
\end{figure} 

The correlation dimension $d_2$ is estimated using the routines included in \cite{hegger1999practical}, based on the Grassberger-Procaccia algorithm  \cite{kantz2004schreiber}. Convergence tests were performed over the input parameters of the algorithm: the embedding quantities $m$, the time delay, and the Theiler window. The Theiler window accounts for the possible oversampling. In fact, the computation of $d_2$ is based on a correlation sum: in the time-series, successive elements are generally not independent and can be highly correlated. To limit this effect, that could lead to inconsistent results, a time-shift -- \emph{i.e.} the Theiler window -- is introduced to reduce the correlation between points during the pair counting.

A limit characterises the computation of the correlation dimension; in particular, the value of the correlation dimension $d_2$ over a decade cannot exceed $d_2=2log_{10}N$, where $N$ is the number of points in the time series (Eckmann-Ruelle limit, \cite{eckmann1992fundamental}). The amount of data available allows us to get a value of $d_{max}\approx 11$. This means that when we approach this upper bound, it is not legitimate to conclude that the $d_2$ value corresponds to the dimension of the inner dynamics.

\medskip
In Fig.~\ref{fig:dimension}, the results are shown with squared-dotted curves. For each positions of the available data-series, we analysed two subsequent blocks of data containing a total of $N=10^5$ points; thus, only values of correlation dimension $d_2<10$ are considered relevant (a green shadowed area indicates the limit). The final value of $d_2$ is obtained by averaging the results of the blocks. We observe that the time-delay $\Delta t$ imposed for the embedding procedure does not influence the final result. A Theiler windows of $\Delta t_{TW}\approx 0.4$ was chosen; we do not observe relevant changes when increasing this parameter. 

For the series \emph{TS-Raw}, we observe a behaviour similar to $m$, with $7 \le d_2 \le 9$ in the span $1.25\le x/D \le 5.75$. For the \emph{TS-POD}, consistent with the previous estimate of $m$, we found that the value of $d_2$ are slightly higher. In the upstream region $x/D\approx 2$, the value close to the upper bound $d_2\approx 10$ does not allow us to conclude that this is the effective value of the correlation dimension. However, the dimension $d_2$ with respect to the embedding dimension $m$ is bounded within the limits imposed by the Takens theorem. 

A decrease is observed at the location extrapolated in the vicinity of the nozzle $x/D<1$ and far downstream, for locations $x/D>6$. Also in this case, low values observed in the \emph{TS-POD} data are due to the POD filtering suppressing important dynamics in these regions of the flow (see \cite{breakey2013near}).

\subsubsection{A brief discussion on the estimated dimensions}\label{sec:discuemb}
\begin{figure}
\centering 
\subfigure[Phase space, $x/D=2.5$]{
\includegraphics[width=0.4\textwidth]{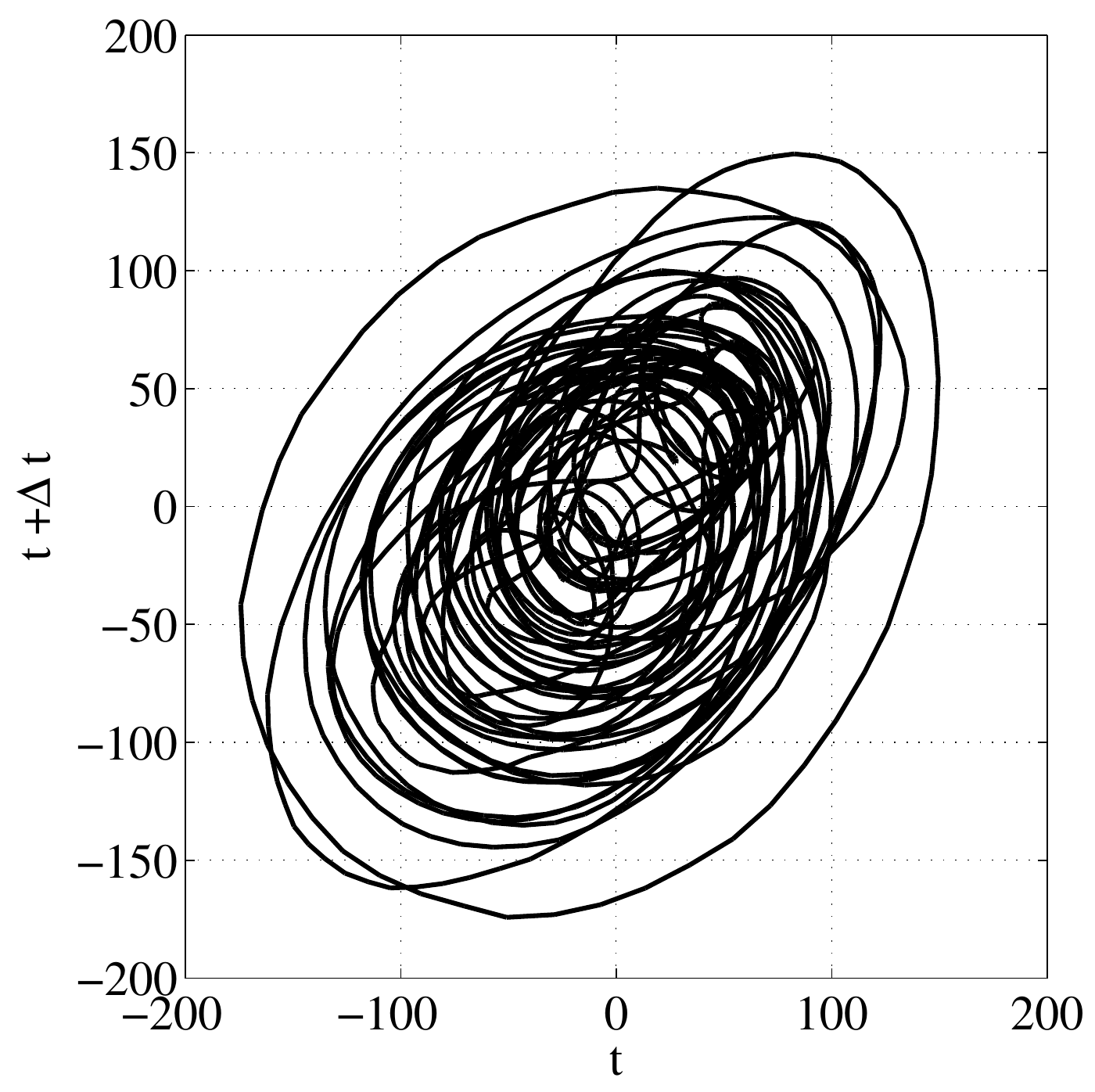}}
\subfigure[Phase space, $x/D=4.1$]{
\includegraphics[width=0.4\textwidth]{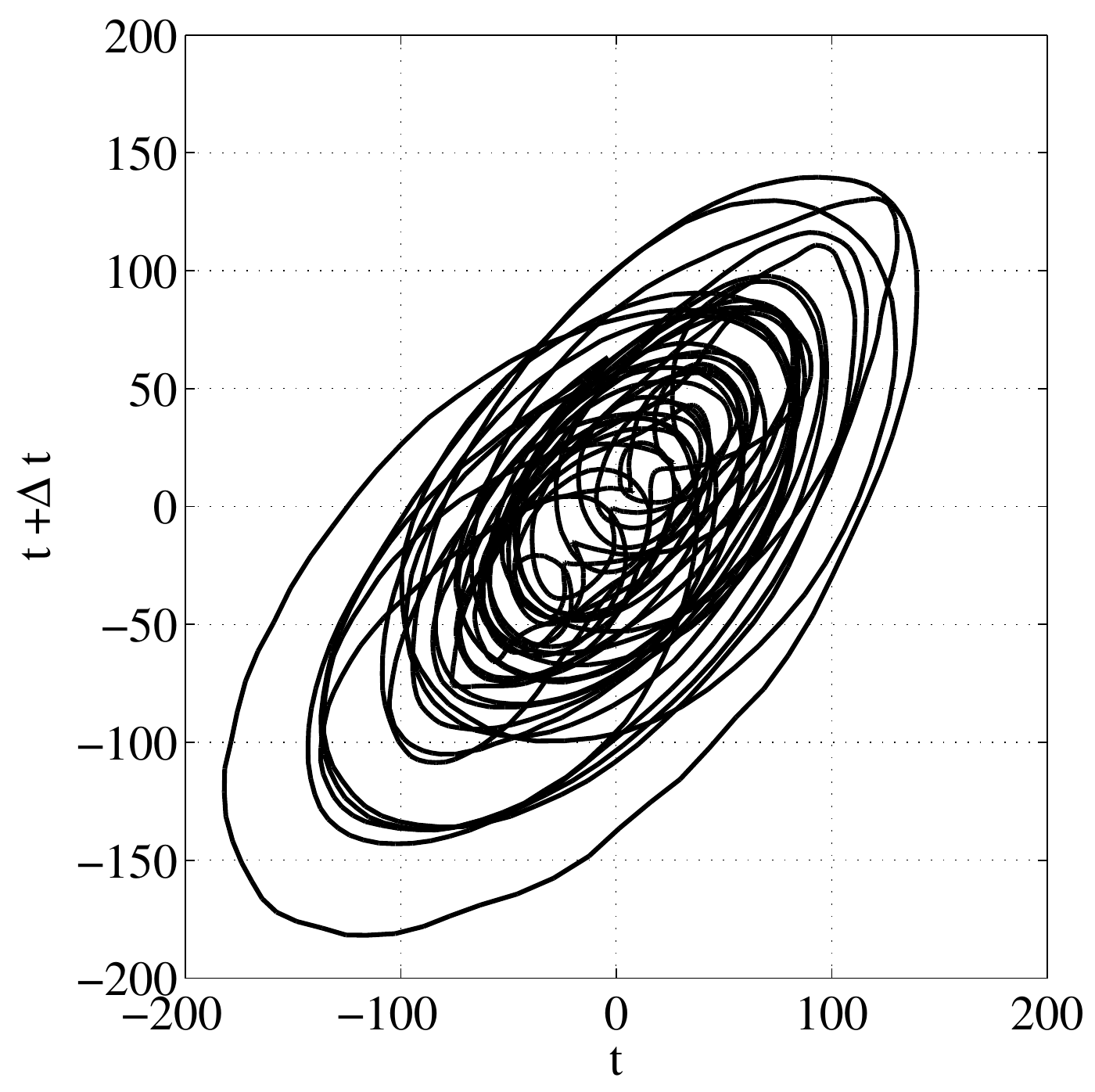}}
\caption{Phase-space analysis for two positions taken from the \emph{TS-POD} series. The embedding space is obtained by time-delayed coordinates, using $\Delta t \approx 0.4$.}
\label{fig:phspa1}
\end{figure} 

The preliminary analysis of the minimal embedding dimension $m$ and the correlation dimension $d_2$ indicates deterministic behaviour, despite the limitations that both algorithms pose. Interestingly, the values of $m$ and $d_2$ are higher when considering the POD-filtered data. This result has been found also by computing the embedding dimension with a different algorithm based on the \emph{false nearest} strategy (results are not reported). We believe that the behaviour is due to the physics of the system described in the different datasets and it is not a numerical artefact. As mentioned in Sec.~\ref{sec:expdata}, the simultaneous pressure measurements of \emph{TS-Raw} are projected on a basis of PODs obtained using two-point flow statistics on a larger number of points in the streamwise direction leading to a finer resolution. Although from the spectral point of view the resulting time-series are equivalent, except at higher $St$ (see Fig.~\ref{fig:tseries1}), from the dynamical point of view the projection on the POD-basis might lead to a richer dynamics due to the increased accuracy of the measurements. When considering the dimensions outside the range  $1.25\le x/D\le5.75$, \emph{i.e.} where the simultaneous measurements are not available, we observe a rather low correlation dimension, while the embedding dimension is constant until $x/D\approx 8$. This effect is due to the extrapolation of the data in this region of the flow. For this reason, in the following sections we focus only on the points $x/D<5.75$.

%
%
\begin{figure}
\centering 
\subfigure[Phase space, $x/D=5.7$]{
\includegraphics[width=0.4\textwidth]{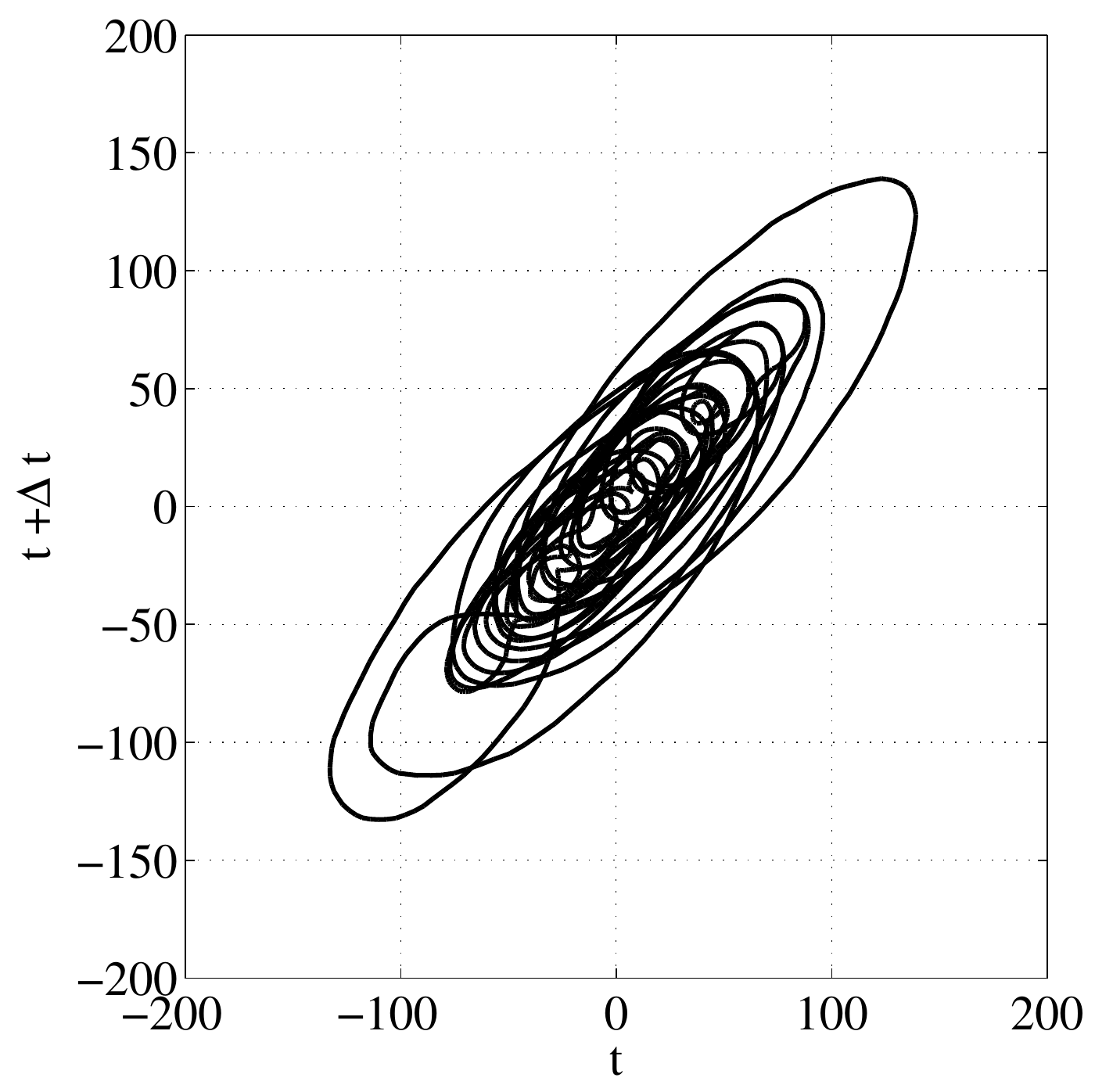}}
\subfigure[First return map, $x/D=5.7$]{
\includegraphics[width=0.4\textwidth]{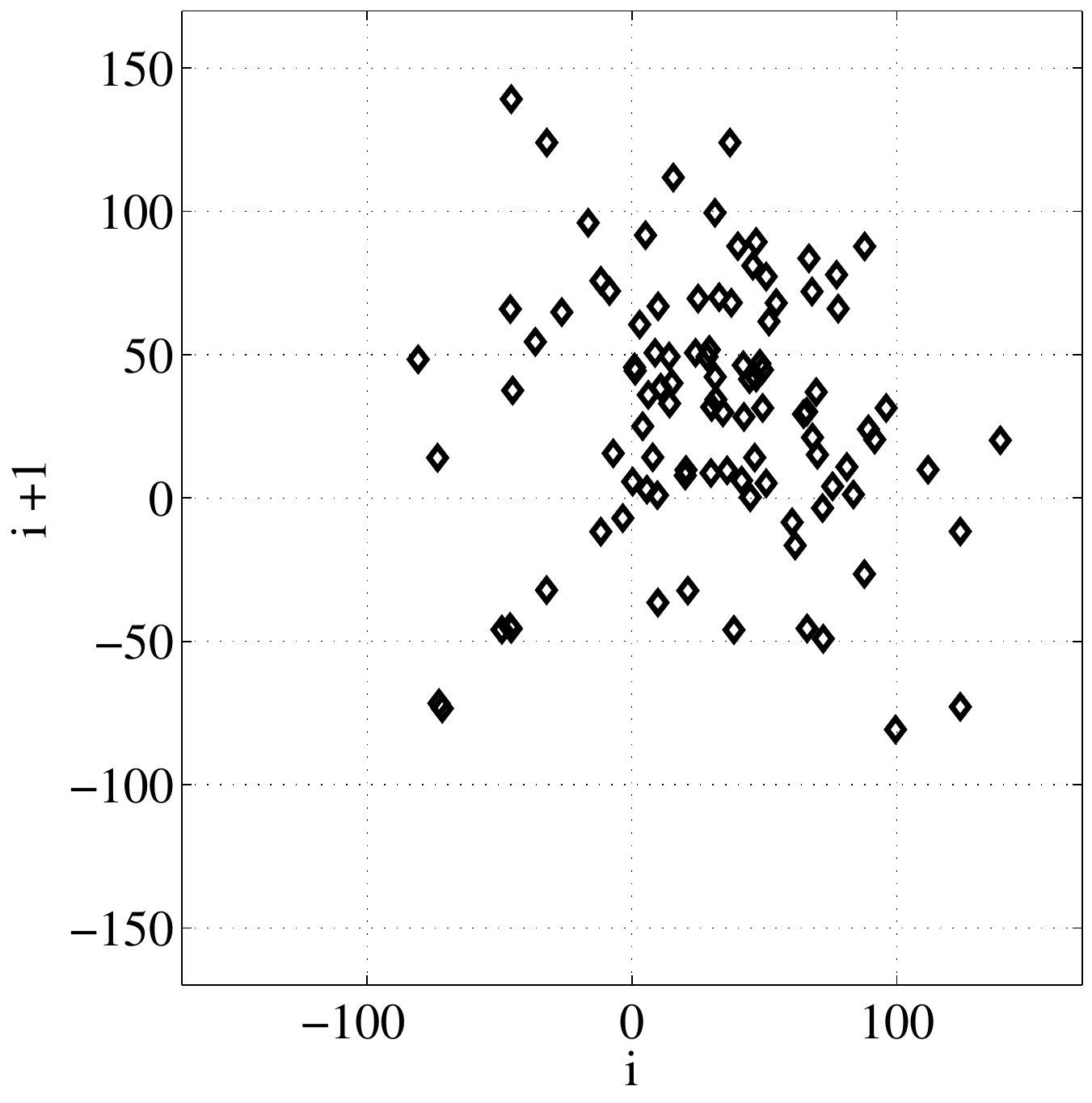}}
\caption{Phase-space analysis for the \emph{TS-POD} series at $x/D=5.7$ (left) and first return map obtained by stacking the local peaks of the time series. The embedding space is obtained by time-delayed coordinates, using $\Delta t \approx 0.4$.}\label{fig:phspa2}
\end{figure} 

\medskip
The embedding space that should be used according to the Takens’ criterion is rather large. In Fig.~\ref{fig:phspa1} and Fig.~\ref{fig:phspa2}$a$, we show three phase-portraits reconstructed in a three dimensional embedding space obtained by imposing a delay of $\Delta t\approx 0.4$, between each of the three coordinates. In all cases, at a first glance, the reconstructed phase space suggests a toroidal nature for the dynamical system; unfortunately, the projection is poor due to the high-dimensionality and cannot be further investigated. Moreover, the first return map in Fig.~$\ref{fig:phspa2}b$ suggests contamination of the dynamics due to external forcings. Similar results were obtained for all the positions along $x/D$ regardless of which embedding strategy was used (SVD embedding and derivative embedding).

\medskip
In conclusion, the results are consistent with a determinism of the time-series under investigation and suggest that the dynamics is relatively low-dimensional. However, the embedding and correlation dimensions are still too large for useful characterisation of the attractor geometry. For this reason, alternative strategies are necessary for exploration of the dynamics of the system. In the next section, system identification is applied: with the same tools, we attempt to replace the original data with surrogate models that capture a part of the dynamics.

%
%
\section{Part II: A system-identification modelling approach}
\label{sec:sysID}

In the previous section, the preliminary study of the time-series -- based on statistical tools and Fourier analysis -- provided clues that the underlying temporal dynamics of the axisymmetric coherent structures is deterministic. However, the dynamics educed locally is rather rich and cannot be analysed by simply using the standard tools from dynamical system theory. From a global view point, the turbulent jet is characterised by a strong convective behaviour: the intrinsic dynamics is sensitive to external forcing \cite{huerre1990local}. Thus, one might wonder to what extent the behaviour educed from the time-series analysis is due to intrinsic non-linear interactions, external activation of degrees of freedom driven by external (stochastic) forcings or a combination of intrinsic and extrinsic mechanisms.

In an effort to understand the underlying dynamics of the system, surrogate models can be introduced by applying system identification \cite{kantz2004schreiber, aguirre2009modeling}. By definition, system identification aims at \emph{building mathematical representations of dynamical systems from set of observables, i.e. measured data}. In other words, given a dynamical system, we can analyse it from its observables and infer a model reproducing the dynamics (or part of it). Starting from the hypothesis that the chosen observables properly represent the system behaviour, this approach can be summarised in the following steps:
\begin{enumerate}
\item \emph{Identification of the model structure and parameter estimation} -- models are identified such that the dynamics of the observables is reproduced. Due to the number of parameters involved, the number of possible models is typically large (see Sec~.\ref{subsec:sysID} and \ref{sec:appA}).
\item \emph{Model validation} -- among all the possible models, only a few allow a proper characterisation of the dynamical system. Validation is necessary in order to discard the models unable to reproduce a dynamics close to the real one.
\item \emph{Prediction or dynamical analysis} -- according to the goal of the identification process, we can aim at minimising the prediction error given a time horizon or analysing the dynamical properties of the system. 
\end{enumerate}
The last aspect is particularly relevant: according to the goal of the identification process, the choices of the inputs/outputs as well as the model structure are of great importance.

 In the following, we will often refer to \emph{noise} indicating with this term the external forcings driving the system dynamics. As already said, turbulent jets are convectively unstable, so strongly sensitive to external forcings; \emph{measurement noise} will be generally indicated as \emph{error} or \emph{residual} in the context of the identification problem.

\subsection{Choice of the model, inputs and outputs}\label{subsec:sysID}
 For our application, the chosen model consists of non-linear Volterra series of polynomials. Hereafter, the outputs will be denoted by ${y}(t)$ and will be represented by the measurements taken at different locations along the streamwise direction. By including also the inputs, indicated by ${u}(t)$, the model reads 
\begin{eqnarray}\label{eq:fullnarmax}
y(t) &=& \sum_{i=1}^{n_{\alpha}} \alpha_i y(t-i) +  \sum_{i=1}^{{n_{\beta}}} \beta_i u(t-i) + NL(u(t),y(t)) +e(t).
\end{eqnarray}

This relation is called the \emph{equation error model}. On the right-hand side, the first term relating the past outputs at the present $y(t)$ is referred to as the \emph{auto-regressive} term; the second term reproduces the dynamics between the inputs $u(t)$ and the outputs $y(t)$ (\emph{exogenous} part). The unknowns of the model are the set of coefficients $\boldsymbol{\alpha}$, $\boldsymbol{\beta}$, the non-linear part -- here represented by high-order polynomials -- and the error $e(t)$.  A classical way to model the error $e(t)$ is to consider it as a \emph{moving-average} of unknown white-noise. The most general strategy of identification based on Eq.~\ref{eq:fullnarmax} is the non-linear (N) autoregressive (AR) exogenous (X) with moving average (MA) algorithm, usually referred as N-ARMAX.

A quick overview of the algorithm is provided in \ref{sec:appA}, while for a deeper discussion we refer to \cite{ljung1999system, guidorzi2003multivariable, semeraro2015techrep}. For our purposes, it is interesting to focus here on the choice of the inputs; in particular,
\begin{enumerate}
\item if we are interested in revealing the core mechanisms of a dynamical system, we can neglect the input $u(t)$ such that all the dynamics is described by the autoregressive part of the model and it is not contained in the inputs; in this case, the model is a discrete representation of an autonomous system.
\item if a non-autonomous representation is sought, an optimal choice is represented by inputs modelled as stochastic white noise or random binary sequences. In this case, all the dynamics will be described by the identified input-output model.
\item a third choice consists in introducing as input $u(t)$ one of the available outputs $y(t)$, thus containing already some of the dynamics. In this case, due to the correlation between input and output, the resulting model will be an optimal transfer function among the elements; this approach is control-oriented.
\end{enumerate}
In sec.~\ref{sec:rom1}, we use full non-linear ARMA models; the main goal is to identify the \emph{essential} dynamics using autonomous models, in order to educe the intrinsic mechanisms at work. In sec.~\ref{sec:rom2}, we aim at identifying models that minimise the prediction error in terms of best mean-squared errors over a time-horizon, given a measurement as input $u(t)$; in this case, full non-linear ARMAX models are computed.

The models discussed in the next sections were computed using the tools developed by Luis~A.~Aguirre and co-workers (see \cite{aguirre2009modeling}). The linear models were cross-validated with in-house scripts \cite{semeraro2015techrep}. We use polynomial expansions, but alternative approaches are available. For instance, non-linear combinations of elementary functions can be used. In this case, the system identification problem changes: we do not seek only the coefficients of the model but also the optimal combination of functions fitting the original data; due to the necessity of identifying also the basis of functions, these techniques require full non-linear optimisation \cite{nelles2001nonlinear} or statistical-learning techniques (see \cite{SINDy}, for instance).

%
%
\section{Non-linear models: dynamical system analysis from the observables}\label{sec:rom1}
In this section we consider non-linear ARMA modelling. With respect to the full equation error model in Eq.~\ref{eq:fullnarmax}, we neglect the exogenous terms. The basic model reads
\begin{eqnarray}
y(t) &=& \sum_{i=1}^{n_{\alpha}} \alpha_i y(t-i) + NL(y(t)) +e(t).
\end{eqnarray}
The identification process consists in identifying the number of necessary terms and coefficients $\boldsymbol{\alpha}$ for an optimal representation of the time-series over a window of assimilation.

This approach leads to a discrete, autonomous model. The un-modelled part is accounted for in the error term $e(t)$. The final analysis is performed on a surrogate time-series obtained by running the model over long time-horizons, not driven from external known forcings. Thus, we do not expect a predictive model, but rather a \emph{phenomenological} model. The aim is to mimic a part of the dynamics of the original system.

We analyse the results obtained by applying the identification approach to the time-series. In particular, we consider the \emph{TS-POD} at $x/D=5.7$, non-linearly filtered (see Sec.~\ref{sec:filtering}). As previously shown, the \emph{TS-POD} series are less affected from measurement noise, in the range where the simultaneous data from \emph{TS-Series} are available; moreover, the non-linear filter preserves the dynamics across all the frequencies. The chosen location corresponds roughly to the end of the potential core. 

\smallskip
In Sec.~\ref{sec:compuSysId}, we provide a brief description of the parametric analysis performed during the computation of the non-linear ARMA models. Results are shown and discussed in Sec.~\ref{sec:resulSysId} for three different models.

\subsection{Computation of the chosen models and validation}\label{sec:compuSysId}
The analysis involves several parameters, namely: i) the polynomial order; ii) the window of assimilation; iii) the maximum number of coefficients $n_{\alpha}$; iv) the optimisation parameters. 
Due to the non-linear terms and the complexity of the system, we expect multiple minima for the optimisation process. In that sense, ranging a large span of parameters is necessary. In the following we briefly motivate and explain our choices.

\paragraph{Polynomial order -- non-linearities} We consider as maximum non-linear order $3$; the choice is a compromise between numerical and computational issues rising from higher-order models and the physical description of the system. In particular, high-order non-linearities might lead to inconsistencies due to singular entries in the Hankel matrices (see \ref{sec:appA}), used during the identification process.

\paragraph{Assimilation window}
The model can depend from the chosen assimilation window. Indeed, considering the spectrogram in Fig.~\ref{fig:spectrogram}, we already observed that the dynamics is characterised by events at given frequencies corresponding to organised behaviour in time-domain. In principle, one should consider more windows of assimilation, stacked over the entire time-span; in practice, for computational and convergence reasons, we consider shorter time-windows of assimilation roughly corresponding to the average length of the high-amplitude events observed in the spectrogram. We tested windows of length $\tau_{aw}=[20.5,\, 30.9,\, 41.2,\,61.8,\, 82.3]$; in the present manuscript, we consider $\tau_{aw}=41.2$, that provided best results.

\paragraph{Coefficient $n_{\alpha}$} This coefficient dictates the second dimension of the regression matrix (see \ref{sec:appA}), the first being the assimilation window. More importantly, it relates past entries with the actual value $y(t)$ during the auto-regression process. We consider the range $n_{\alpha}=[1,\, 30]$; the maximum value in convective time corresponds to $\tau\approx 1.25$ and it is based on the analysis of the auto-correlation. Larger values would lead to erroneous correlations between past and present dynamics.

\paragraph{Optimisation process parameters} For the optimisation process, we consider a maximum of $N_i=20$ iterations for each model and a maximum $N_e=60$ elements for the resulting models, whose $10$ terms are related to the error part.

\smallskip
As a result of the parametric analysis, we identified a total of $36000$ models. As said, the models can be biased by the window of assimilation. At a first glance, one might think that the choice of relatively short time-windows of assimilation reduces the validity of the models; however, the main aim of the modelling approach is to explore the dynamics behind the organised behaviour observed in Fig.~\ref{fig:spectrogram}, trying to separate the inherent dynamics observable in small windows of observation from the effects of the noise-sensitivity. In that sense, it might not be possible to get endogenous models of the system, unless we were to introduce an exogenous forcing decorrelated from the output $y(t)$. For this reason, we believe that a short window of assimilation results in an identification process consistent with a part of the inherent dynamics of the dynamical system. 

\subsubsection{Validation process}
The simulation of the models is not driven by known forcings; thus, these non-linear models will quickly deviate from the original time-series after a short transient roughly corresponding to the initial conditions. The validation process is performed in two steps: i) the minimisation of the error during the system identification; ii) the spectral analysis of the model, compared with the original time-series. In particular, we can not apply the Welch method due to the short time-window of assimilation; as an alternative, we apply a power spectral density estimation based on the Yule-Walker method \cite{stoica2005spectral}. This method estimates the spectral content by fitting an autoregressive linear prediction filter to the signal. By applying these two criteria and crossing the results, we were able to identify three different ``families'' of models. Of all the models identified, around $21\%$ show periodicity, quasi-periodicity, toroidal features or chaotic behaviour. The remaining models are badly conditioned, unstable or decaying to zero after a transient. The selected models are robust with respect to the initial conditions, \emph{i.e.} they show the same asymptotic behaviour in the presence of uncertainties in the initial conditions. With respect to the position $x/D$, the optimisation procedure leads to models with different coefficients $\boldsymbol{\alpha}$, but similar dynamics from the phenomenological point of view. Thus, the surrogate models discussed here are still able to qualitatively describe the physics of the underlying system.

\subsection{Autonomous models}\label{sec:resulSysId}
\begin{figure}
\centering 
\subfigure[Comparison between time-series and model]{\includegraphics[width=0.575\textwidth]{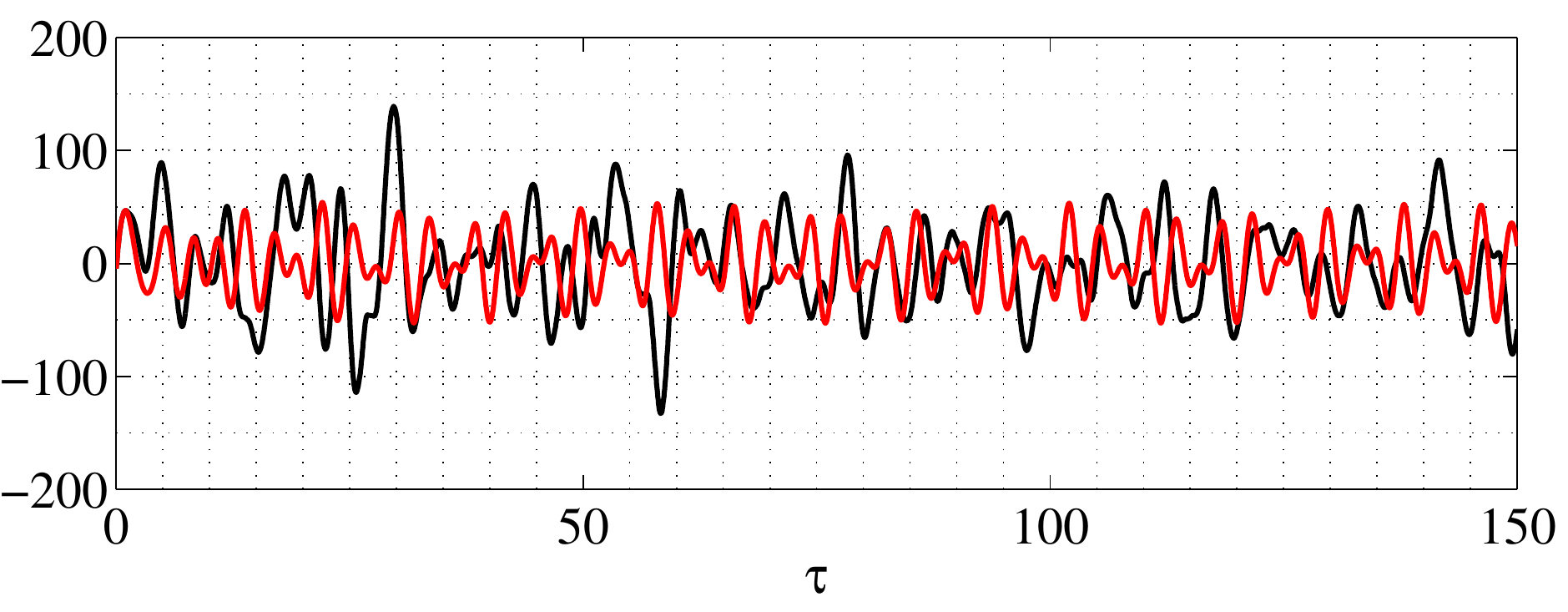}}\\
\subfigure[Autoregressive PSD]{\includegraphics[width=0.3\textwidth]{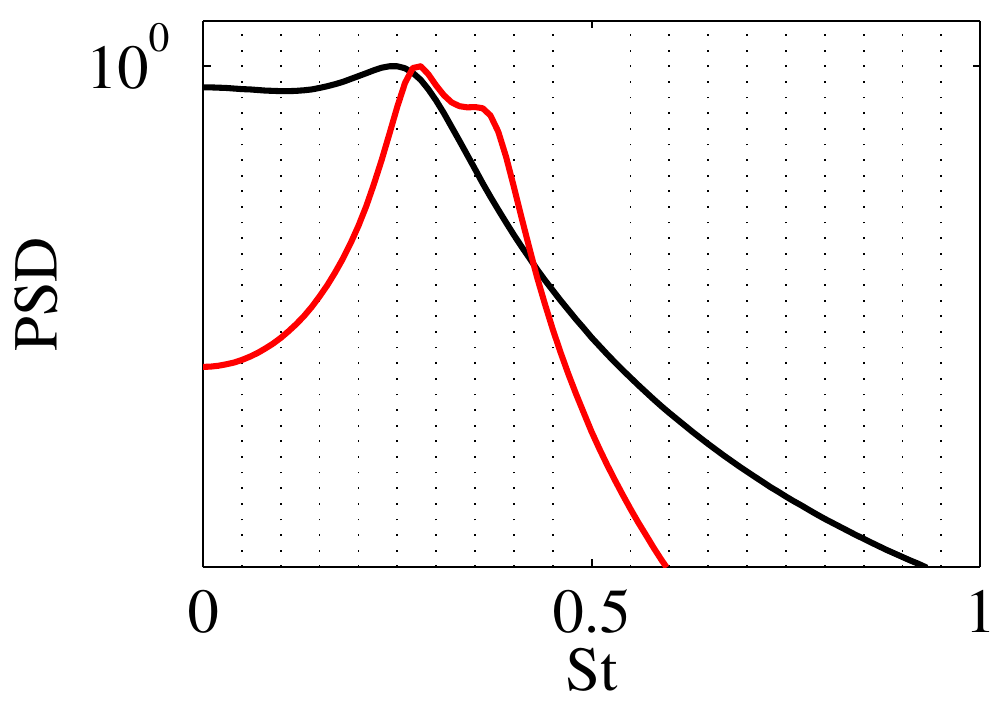}}
\subfigure[Welch PSD]{\includegraphics[width=0.3\textwidth]{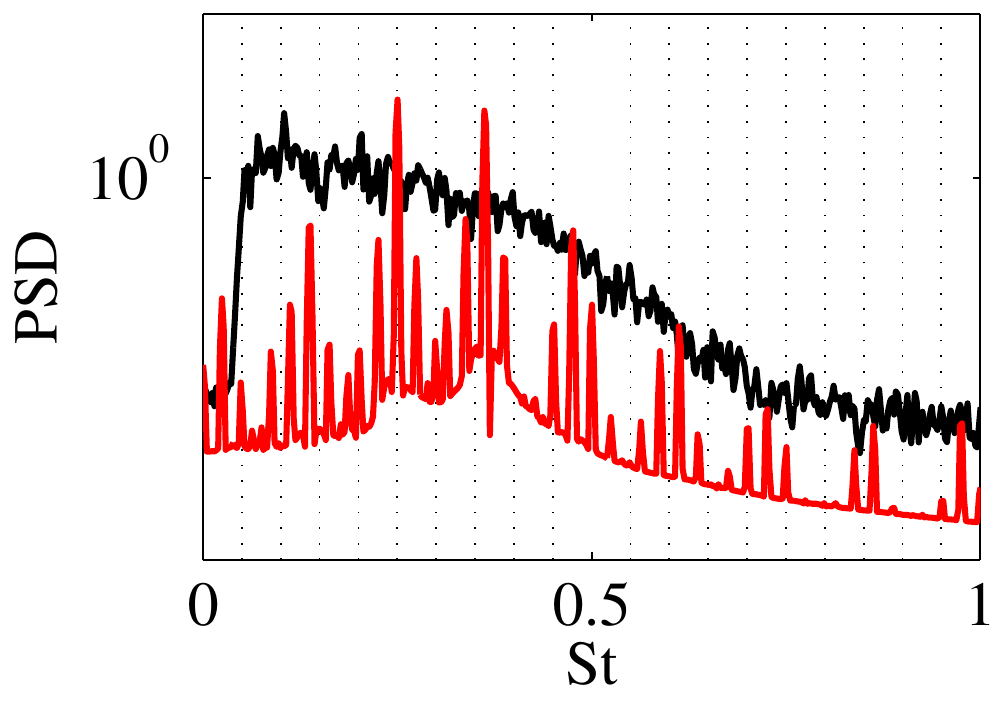}}\\
\subfigure[Phase space comparison]{\includegraphics[width=0.35\textwidth]{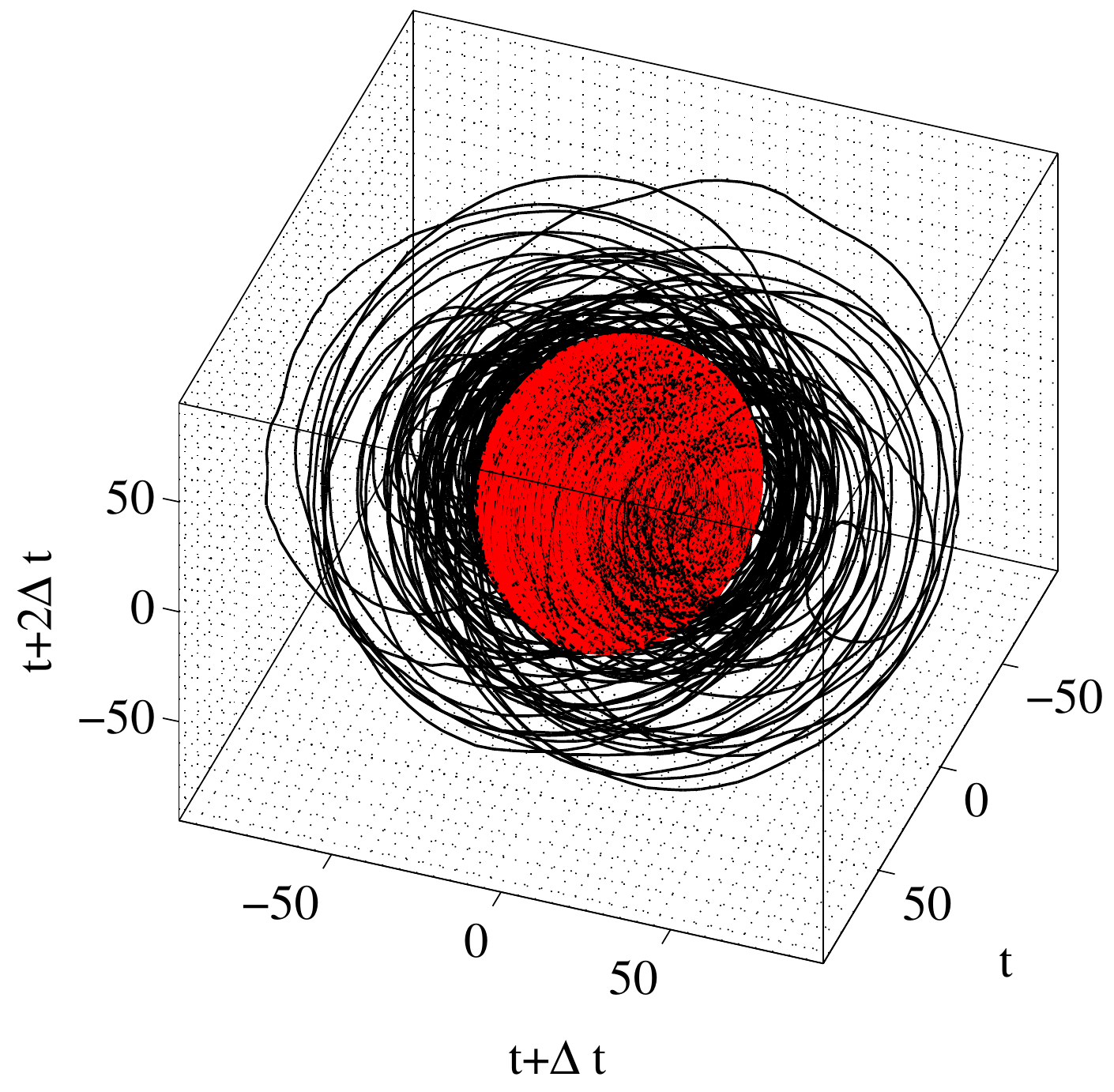}}
\subfigure[First-return map]{\includegraphics[width=0.28\textwidth]{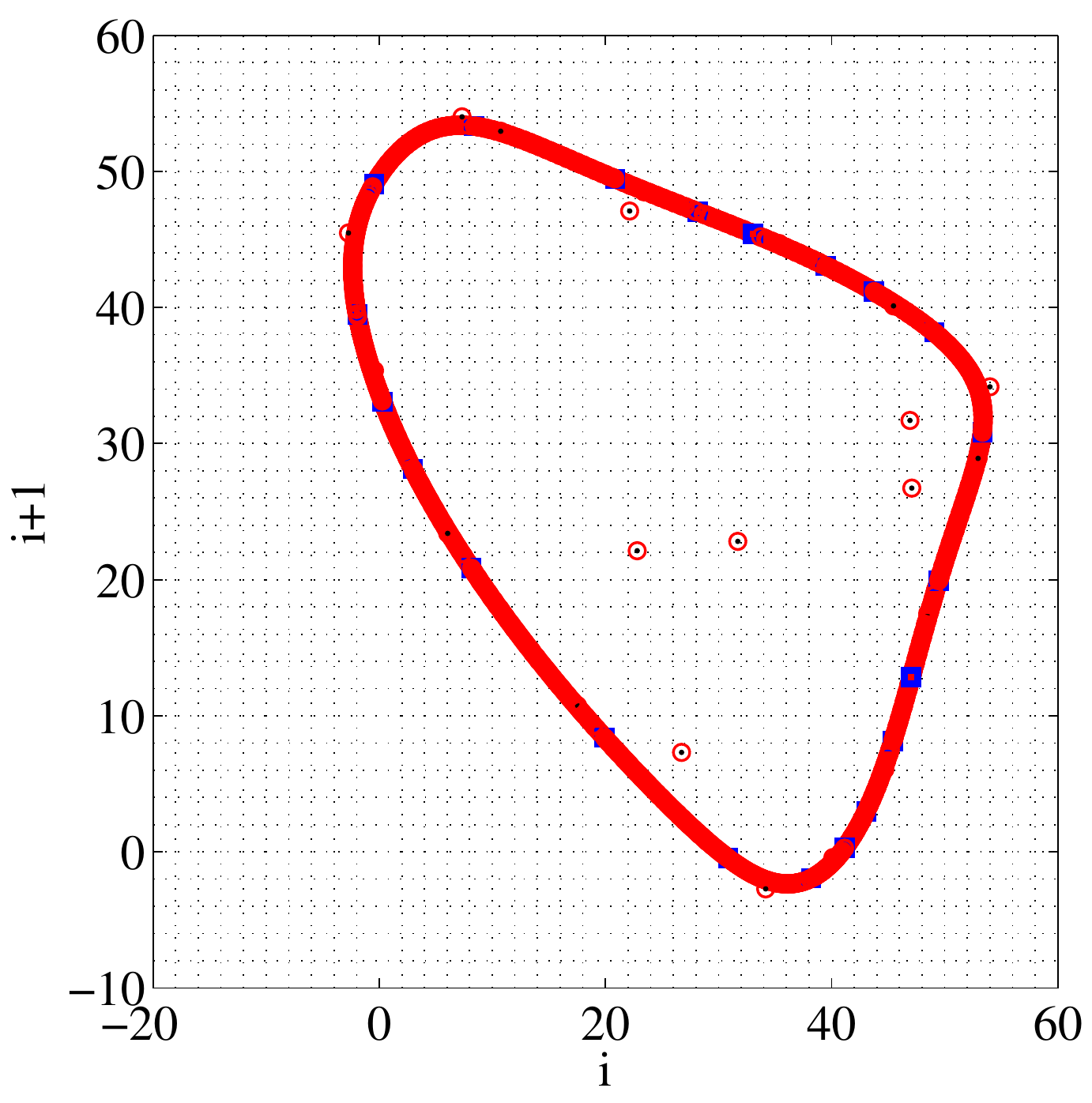}}
\caption{Comparison N-ARMA model $T$ (in red for all the insets) vs. original time-series \emph{TS-POD} at $x/D=5.7$ (in black). The comparison in time-domain $(=a)$ is only qualitative. Thus, spectra analysis $(b-c)$ and phase space reconstruction are necessary for an assessment of the performance of the model. The embedding space for the phase space reconstruction $(d)$ is obtained with a delay $\Delta t\approx 0.3$. The first-return map is obtained using the maximum values in the model $(e)$; the black dots correspond to the beginning of the trajectory, while the superimposed blue dots the last points of the simulation.}
\label{fig:model_arma_1}
\end{figure}

In this section, we discuss three different models, labelled as follows
\begin{enumerate}
\item \emph{Model T} -- in Fig.~\ref{fig:model_arma_1} the model is characterised by $n_{\alpha}=26$. It is a toroidal (T) example, that settles on the frequencies characterising the windows of learning.
\item \emph{Model LC} -- in Fig.~\ref{fig:model_arma_2} the model is characterised by $n_{\alpha}=25$. After a rather long transient (up to $\tau \approx 150$), it ends up on a stable limit-cycle (LC), see Fig.~\ref{fig:model_arma_2_end}. 
\item \emph{Model S} in Fig.~\ref{fig:model_arma_3} the model is characterised by $n_{\alpha}=6$. It is a chaotic model, as it can be seen from the first-return map. We believe that it reproduces some short (S) time dynamics related to the jittering.
\end{enumerate}
The coefficient $n_{\alpha}$ gives to the models distinctive features. The first two models are characterised by a long window of correlation. The third model is obtained considering the shortest, possible correlation length. In the following paragraphs, we comment on the significance of these phenomenological models.

%
%

\subsubsection{Models $T$ and $LC$}
In Fig.~\ref{fig:model_arma_1}, the properties of the model $T$ are analysed. As already mentioned, none of these models is predictive: after a short time, the dynamics quickly deviate from the original one (Fig.~\ref{fig:model_arma_1}$a$). However, the original time-series are contaminated by the external disturbances, while the model behaviour is dictated by the intrinsic dynamic educed during the assimilation process. The quality of the model can be assessed by analysing the spectral properties in the window of assimilation: in Fig.~\ref{fig:model_arma_1}$b$, the power spectral density is shown based on autoregressive modelling; the current model reproduces the peak frequency $St=0.25$; a second peak appears at $St\approx 0.35$. However, we can also observe that model $T$ is not properly reproducing the low frequencies $St\rightarrow 0$. The long-time behaviour is characterized by a toroidal behaviour in phase space, showed with a red solid line in Fig.~\ref{fig:model_arma_1}$d$. The Fourier analysis in Fig.~\ref{fig:model_arma_1}$c$  and the first-return map built with the relative maxima in Fig.~\ref{fig:model_arma_1}$e$ confirm the phase space representation. In particular, in Fig.~\ref{fig:model_arma_1}$e$, the black dots correspond to the beginning of the trajectory, while the superimposed blue dots the final part.

\begin{figure}
\centering 
\subfigure[Comparison between time-series and model]{\includegraphics[width=0.575\textwidth]{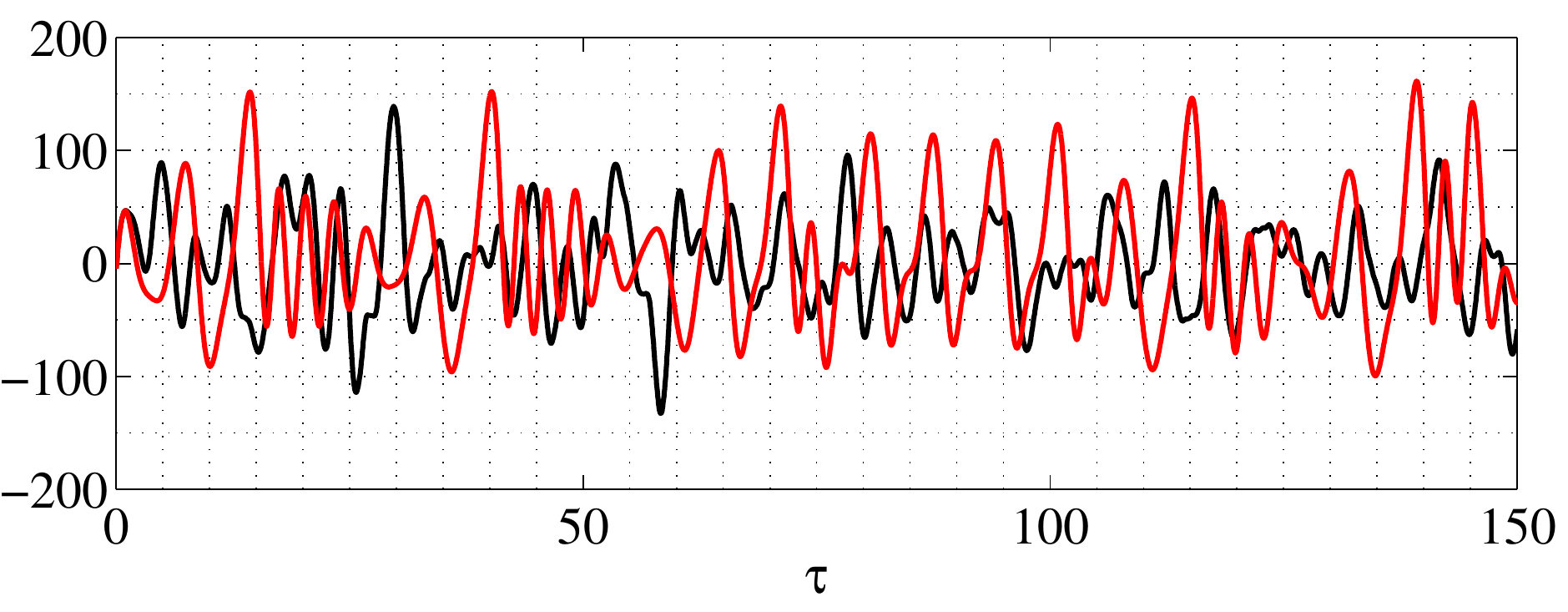}}\\
\subfigure[Autoregressive PSD]{\includegraphics[width=0.3\textwidth]{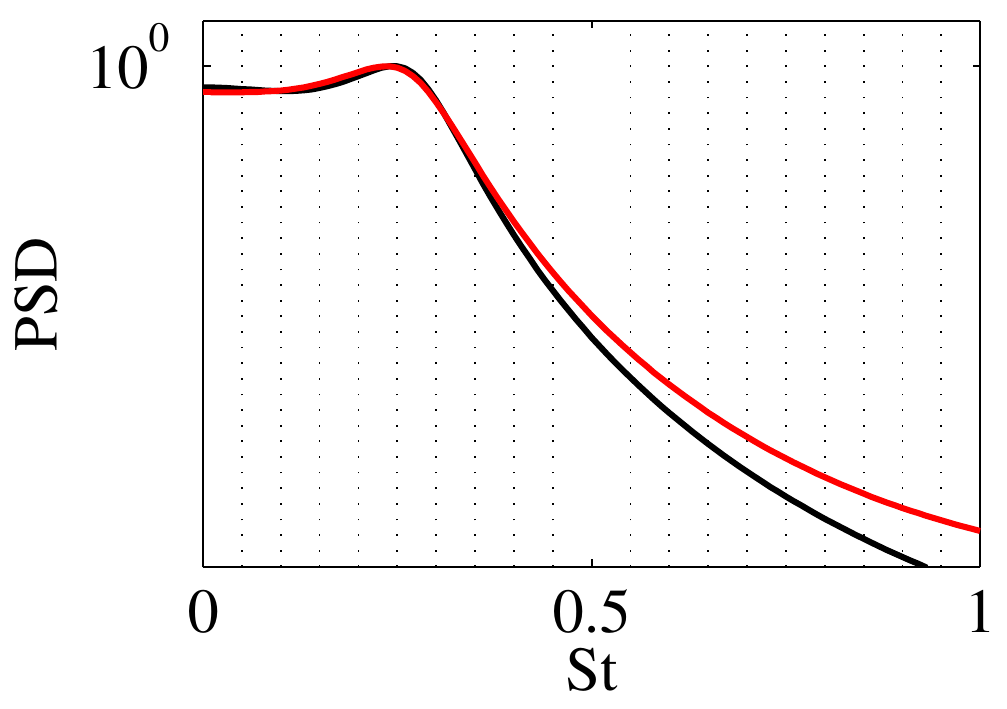}}
\subfigure[Welch PSD]{\includegraphics[width=0.3\textwidth]{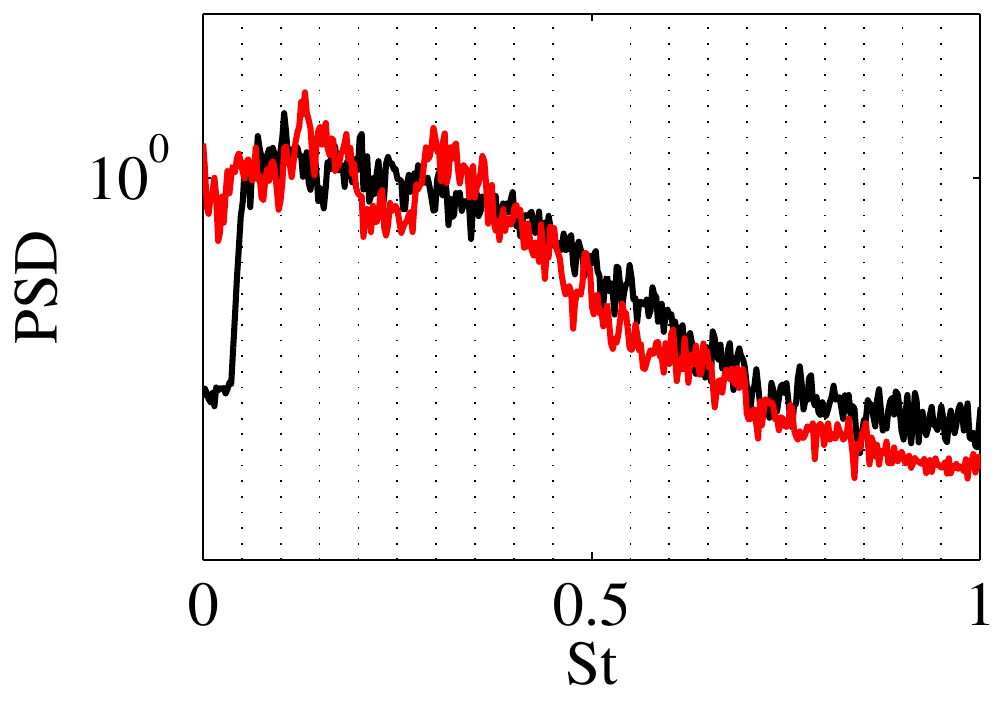}}\\
\subfigure[Phase space comparison]{\includegraphics[width=0.35\textwidth]{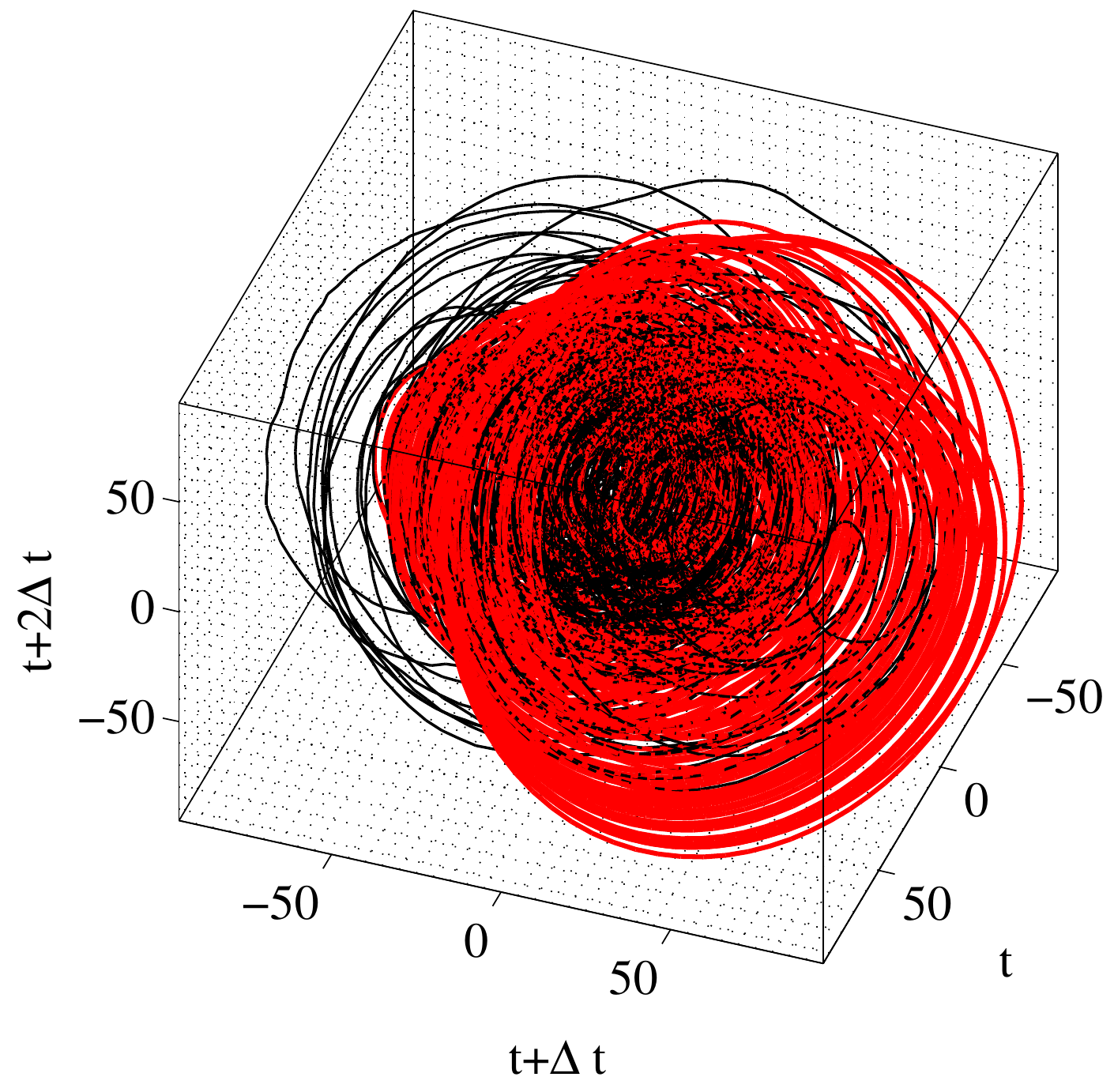}}
\subfigure[First-return map]{\includegraphics[width=0.28\textwidth]{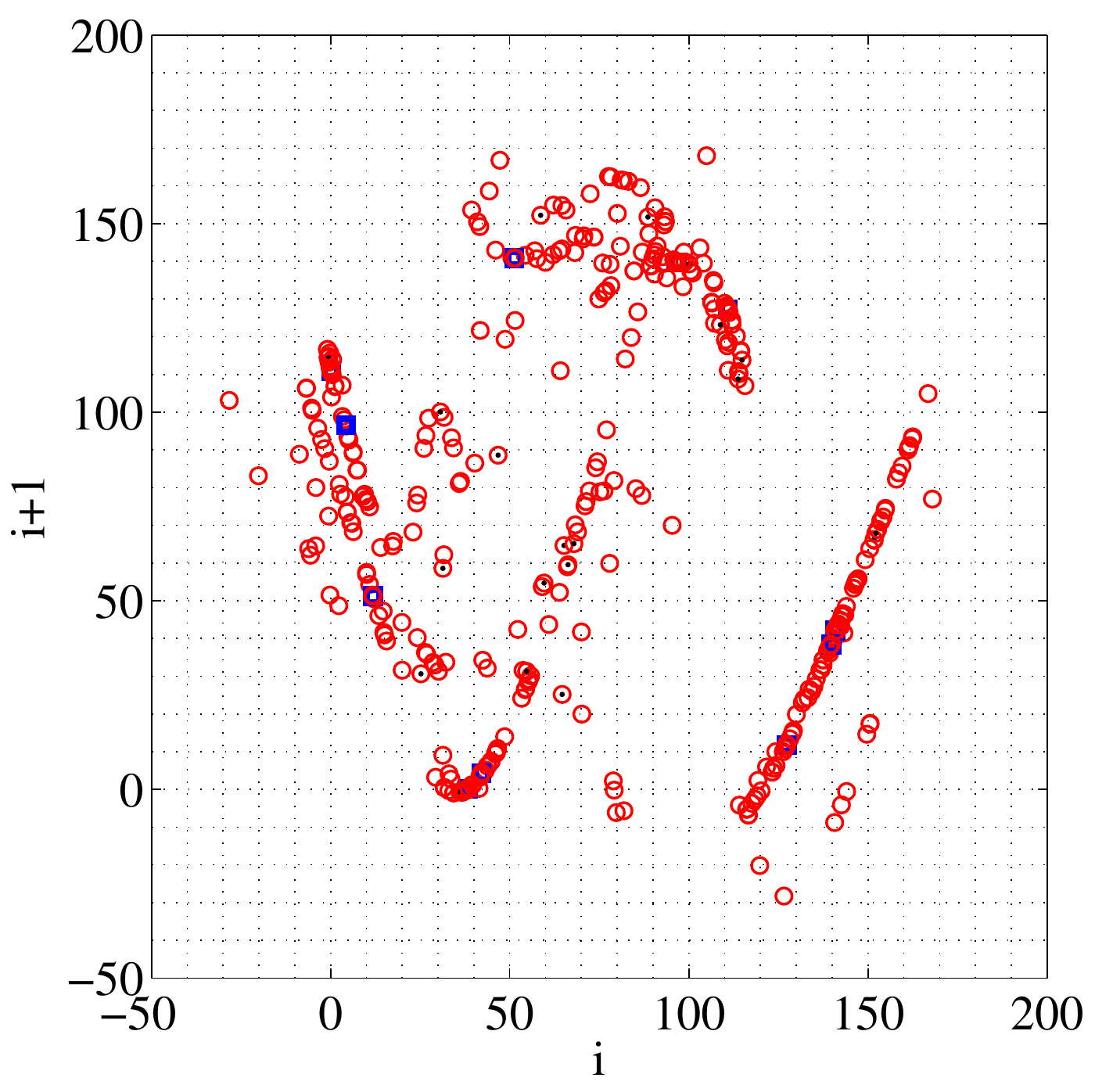}}
\caption{Comparison N-ARMA model $LC$ (in red for all the insets) vs. original time-series \emph{TS-POD} at $x/D=5.7$ (in black). The comparison in time-domain $(=a)$ is only qualitative. Thus, spectra analysis $(b-c)$ and phase space reconstruction are necessary for an assessment of the performance of the model. The embedding space for the phase space reconstruction $(d)$ is obtained with a delay $\Delta t\approx 0.3$. The first-return map is obtained using the maximum values in the model $(e)$; the black dots correspond to the beginning of the trajectory, while the superimposed blue dots the last points of the simulation.}
\label{fig:model_arma_2}
\end{figure}

\begin{figure}
\centering 
\subfigure[Phase space]{\includegraphics[width=0.35\textwidth]{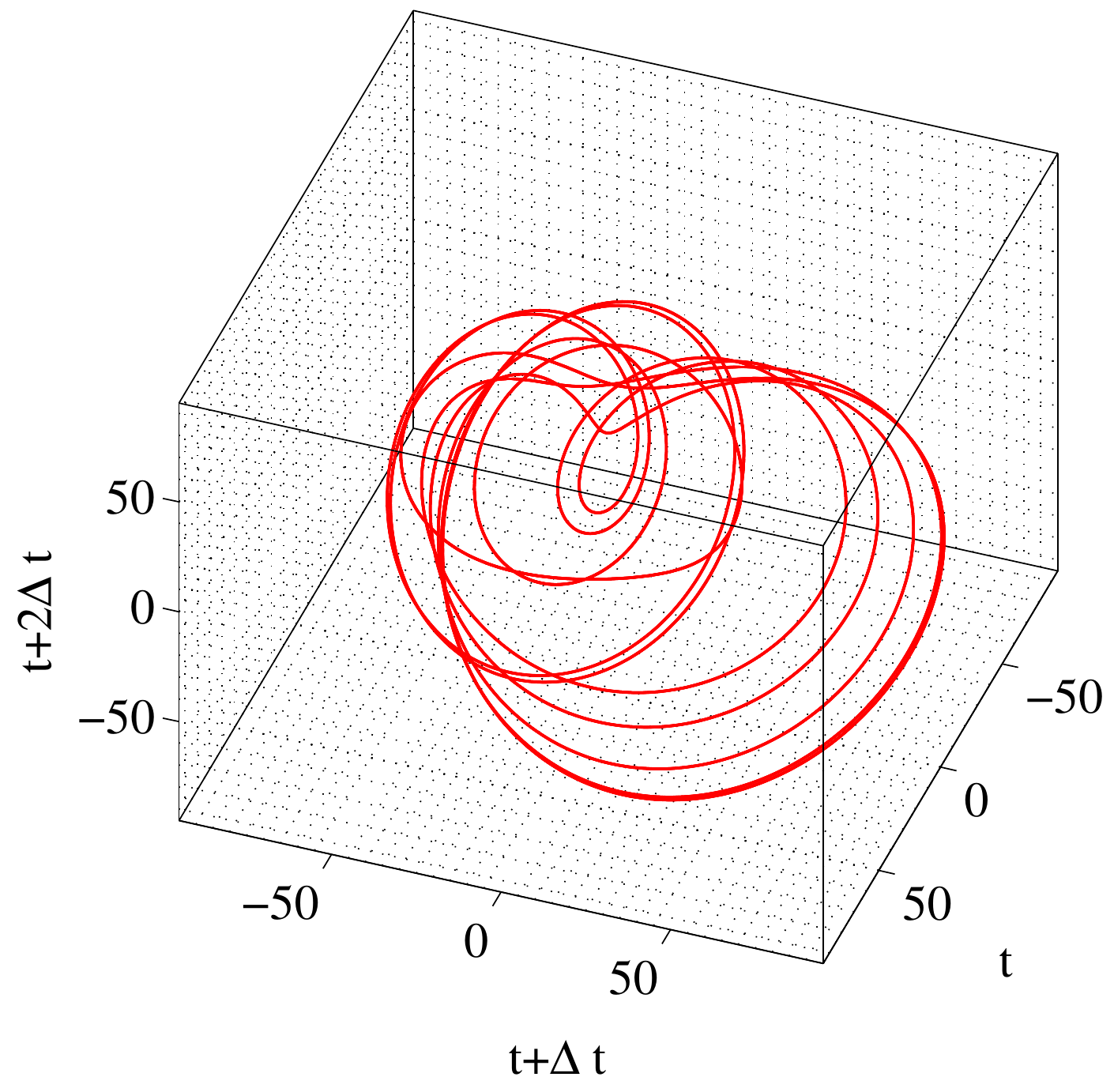}}
\subfigure[First-return map]{\includegraphics[width=0.28\textwidth]{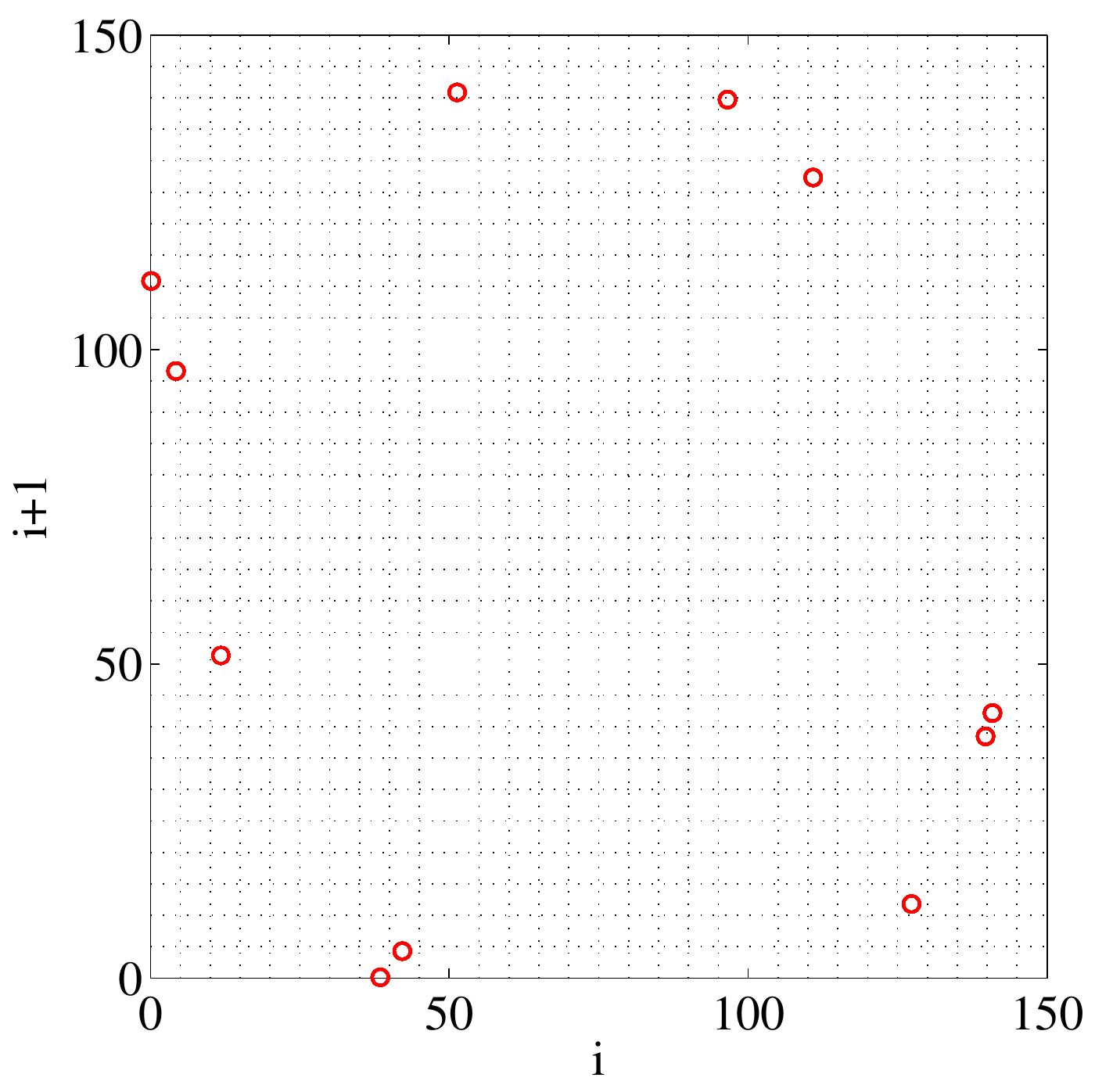}}
\subfigure[Welch PSD]{\includegraphics[width=0.28\textwidth]{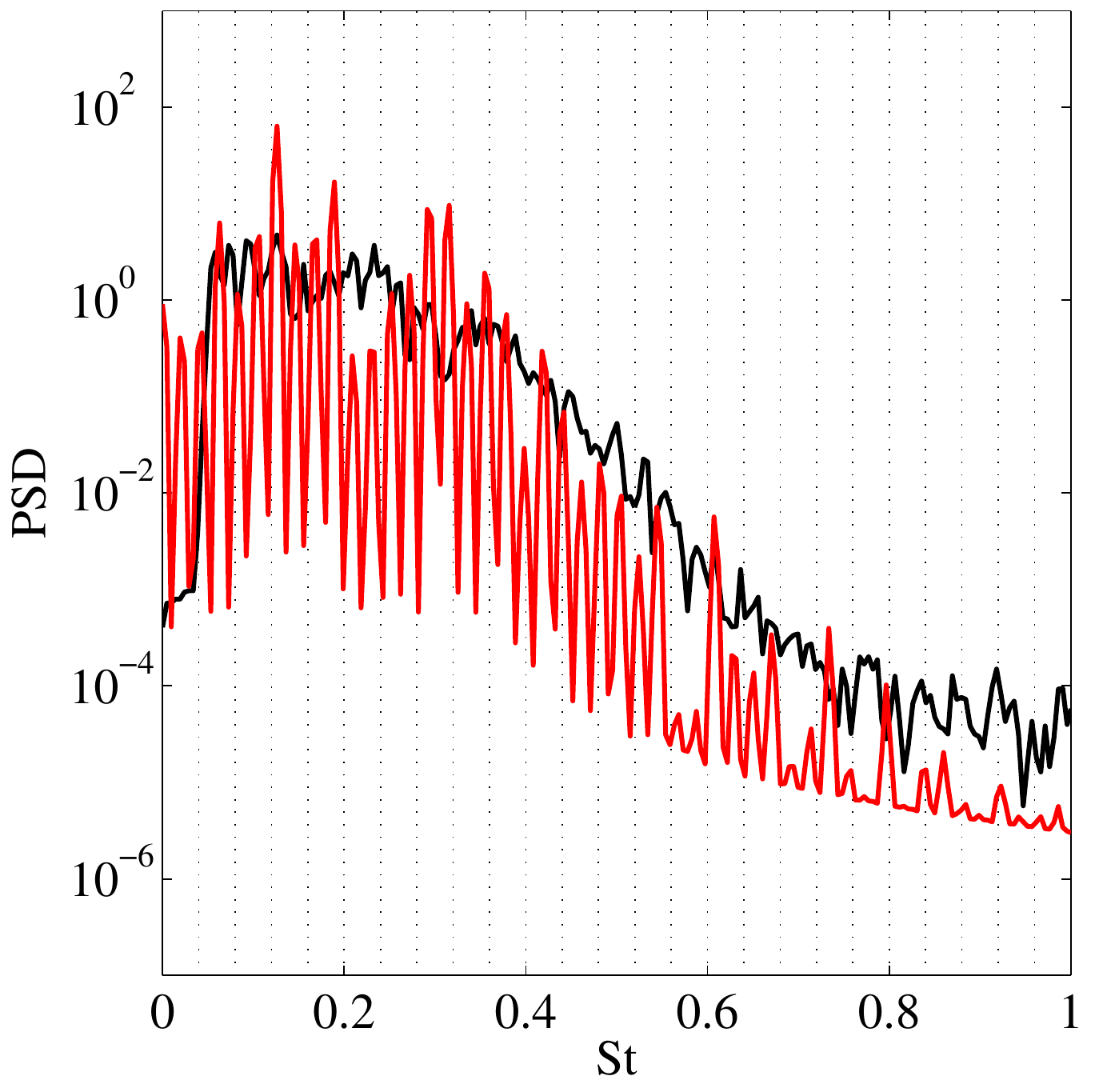}}
\caption{Asymptotic behaviour for N-ARMA model $LC$ shown in Fig.~\ref{fig:model_arma_2}$a-e$: after the transient, the system settles on a stable limit cycle. In $a$ the phase space is shown (embedding delay $\Delta t\approx 0.3$); in $b$ is shown the first-return map (blue dots of Fig.~\ref{fig:model_arma_2}$e$).}
\label{fig:model_arma_2_end}
\end{figure}

\smallskip
Model $T$ is characterised by a limited match in the autoregressive PSD; for this reason, a second model is presented in Fig.~\ref{fig:model_arma_2}, labelled as model $LC$. In this case, we observe an excellent agreement at low Strouhal numbers, $St<0.5$ (see Fig.~\ref{fig:model_arma_2}$b$). We note a rather different behaviour when comparing the transient with the asymptotic state. During the transient, the spectrum is not characterized by distinct frequencies (see Fig.~\ref{fig:model_arma_2}$c$), although we observe the dominance of frequencies around $St=0.35$ and $St=0.15$ (due to the assimilation windows). The overall dynamics is more complex as shown in the phase space portrait in  Fig.~\ref{fig:model_arma_2}$d$; the corresponding first-return map in Fig.~\ref{fig:model_arma_2}$e$ can be understood by considering the long-time behaviour of the system reported in  Fig.~\ref{fig:model_arma_2_end}. Indeed, after a transient of $\tau\approx 150$, the model settles on a limit cycle (Fig.~\ref{fig:model_arma_2_end}$a-b$); the dominating frequencies in Fig.~\ref{fig:model_arma_2_end}$c$ are the same as those observed in the Welch PSD analysed in Fig.~\ref{fig:model_arma_2}$c$. Having the long-time behaviour in mind, it is easier to understand the first-return map in Fig.~\ref{fig:model_arma_2}$e$. In particular, the blue dots are representative of the limit-cycle trajectory: we can observe that the model during the transient progressively settles on the limit cycle, that ``attracts'' the trajectory.

%
%

%
%

\begin{figure}
\centering 
\subfigure[$TS-POD$ at $x/D=5.7$]{\includegraphics[width=0.325\textwidth]{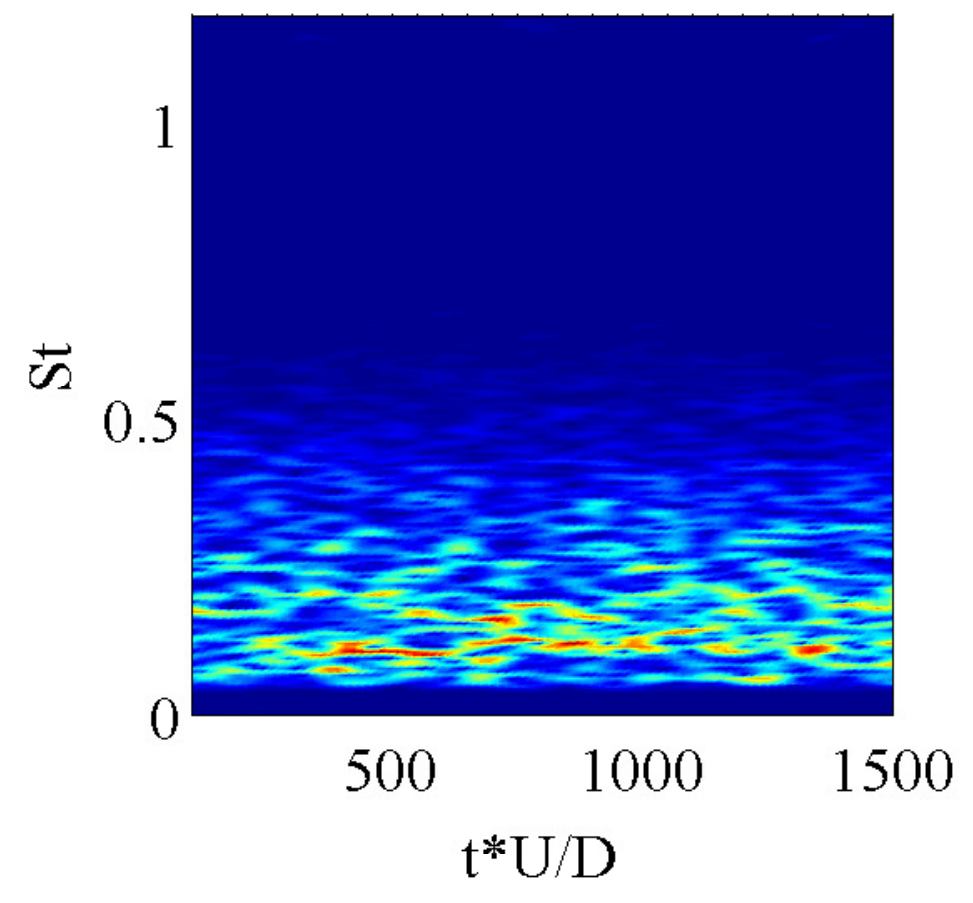}}
\subfigure[Model $2$ at $x/D=5.7$]{\includegraphics[width=0.325\textwidth]{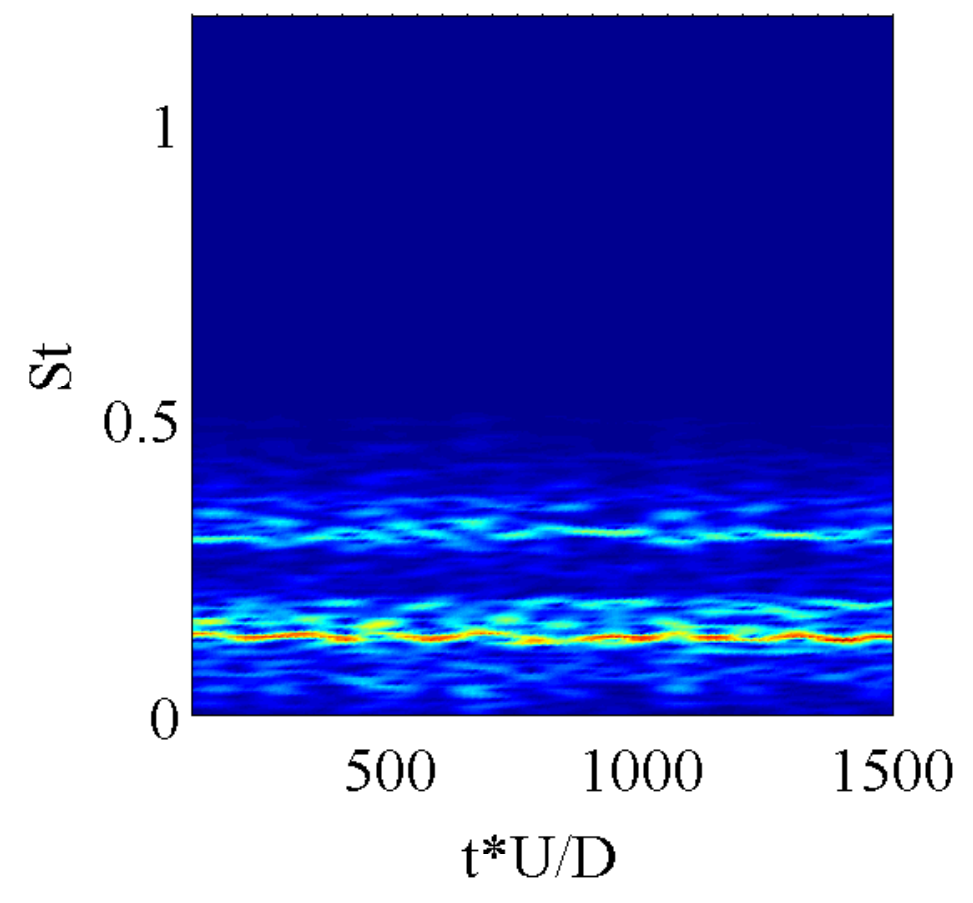}}
\caption{Spectrogram: comparison between the original time-series and the  model $LC$.}
\label{fig:spectromodel}
\end{figure} 

The transient has a qualitative behaviour similar to the original model also if observed from the spectrogram. In Fig.~\ref{fig:spectromodel}, it is possible to compare the original time-series (inset $a$) with the model ($b$), in the same time-span. It is evident that on average the length of the organised events is similar, while a significant difference is given by the frequency range; in particular, as already observed, this feature is an attribute of the extrinsic dynamics at work on the dynamical system. Thus, these two models suggest that if only the endogenous dynamics were considered, the dynamics due to non-linear interactions among wavepackets would be dominated by the organised behaviour that characterises the ``short-living'' events observed in the spectrogram.  In particular, the dynamics would evolve after some transient on the surface of a $2$-torus (T2) or a limit cycle. In models $T-LC$ however, one limit cycle (at least) is stable and attracting. Therefore, the hypothetic underlying chaotic toroidal set is only transient. 
 
\subsubsection{Model S}
\begin{figure}
\centering 
\subfigure[Comparison between time-series and model]{\includegraphics[width=0.575\textwidth]{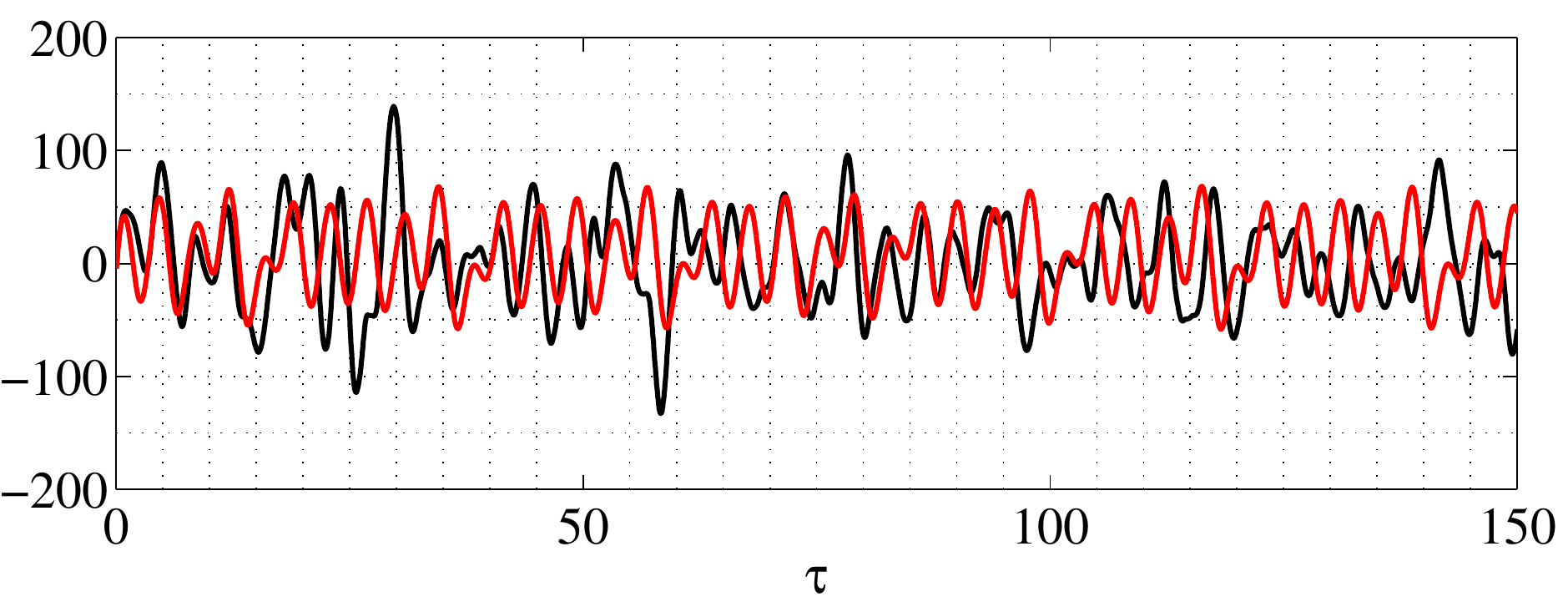}}\\
\subfigure[Autoregressive PSD]{\includegraphics[width=0.3\textwidth]{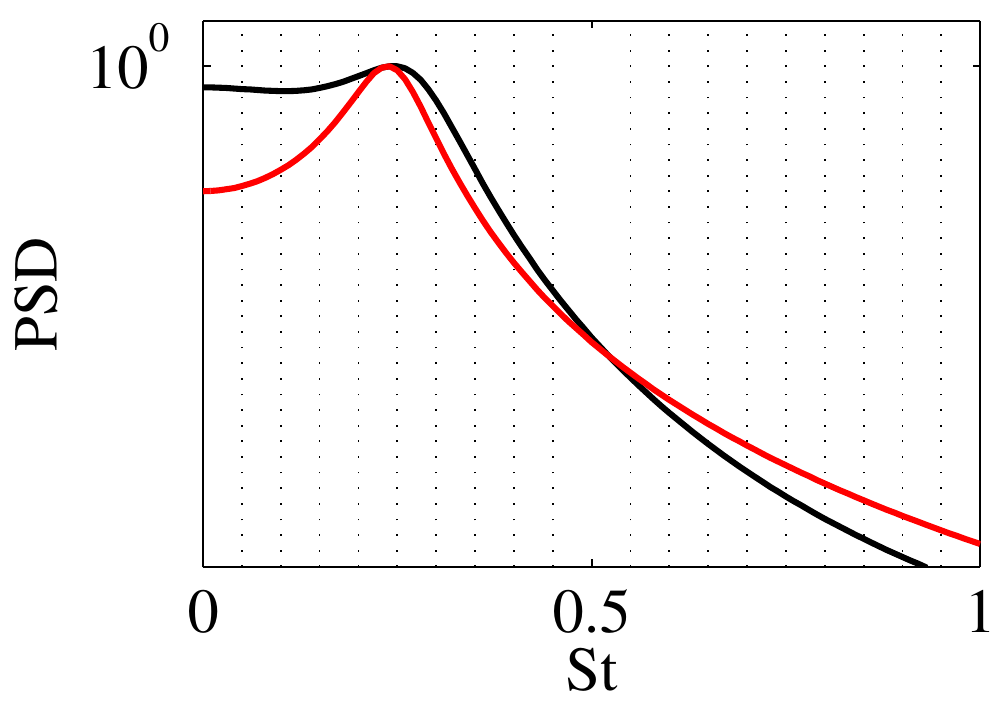}}
\subfigure[Welch PSD]{\includegraphics[width=0.3\textwidth]{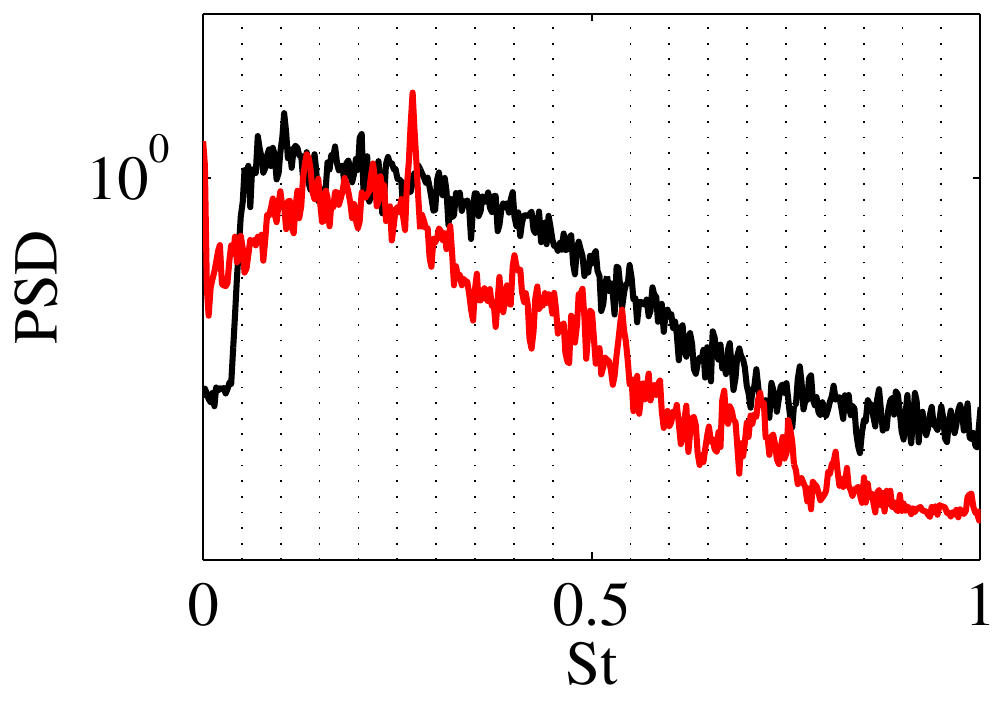}}\\
\subfigure[Phase space comparison]{\includegraphics[width=0.35\textwidth]{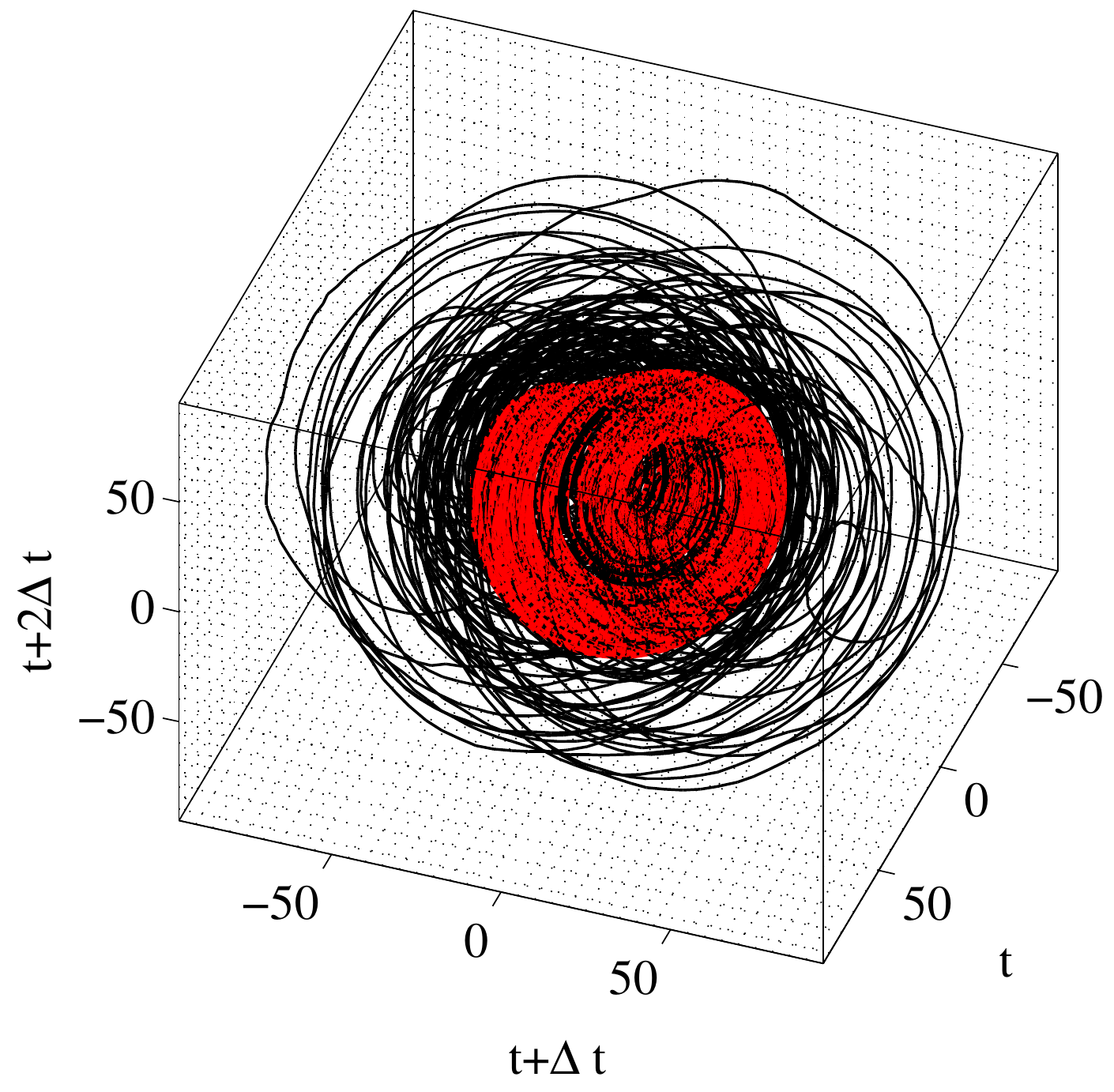}}
\subfigure[First-return map]{\includegraphics[width=0.28\textwidth]{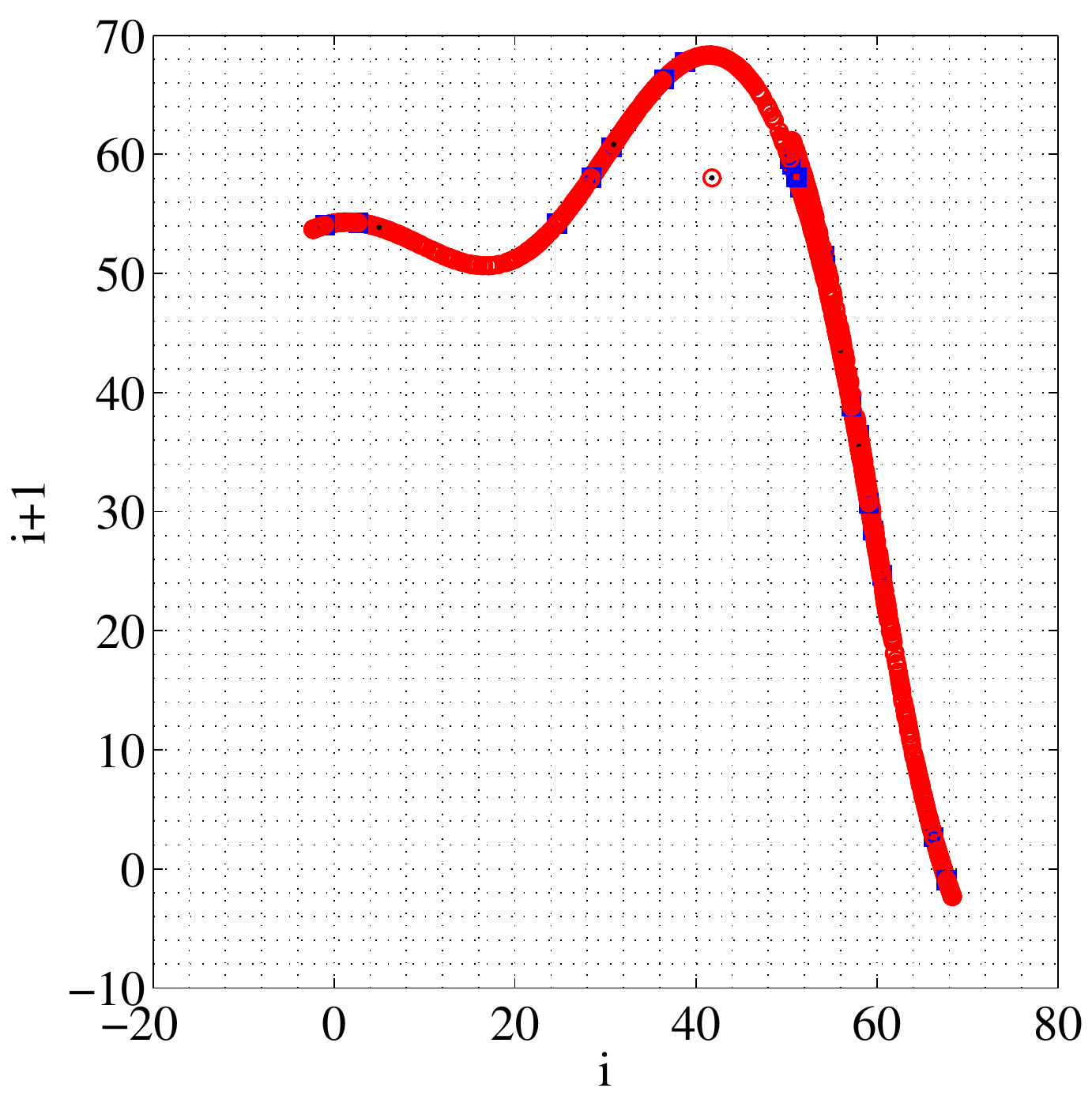}}
\caption{Comparison N-ARMA model $S$ (in red for all the insets) vs. original time-series \emph{TS-POD} at $x/D=5.7$ (in black). The comparison in time-domain $(a)$ is only qualitative. Thus, spectra analysis $(b-c)$ and phase space reconstruction are necessary for an assessment of the performance of the model. The embedding space for the phase space reconstruction $(d)$ is obtained with a delay $\Delta t\approx 0.3$. The first-return map is obtained using the maximum values in the model $(e)$; the black dots correspond to the beginning of the trajectory, while the superimposed blue dots the last points of the simulation.}
\label{fig:model_arma_3}
\end{figure}

In this section, we analyse a third model, depicted in Fig.~\ref{fig:model_arma_3} and based on a shorter window of correlation. Also in this case, the peak in frequency attains at $St\approx 0.25$, as shown in the auto-regressive PSD analysis in Fig.~\ref{fig:model_arma_3}$b$; this same dominance is observed in the Welch analysis in Fig.~\ref{fig:model_arma_3}$c$. The resulting behaviour in phase space and in the first-return map is rather distinctive and different compared to the previous models. In particular, this model appears to slowly evolve in time; in this case, we can conjecture that the model captures some inherent mechanisms related to combination of wavepackets-pairs being driven by the stochastic turbulence in the real case.

\subsubsection{Discussion}
Within the limits of the methodology, an analysis of the models provides some hints regarding the possible dynamics underlying the system. The main idea is that the jittering is a mixture of stochastic and deterministic components. \emph{Model T} and \emph{Model LC}  seems to indicate that frequencies can lock together, leading to non-linear wave-wave interactions. Eventually, without external forcing, they can set on limit cycles or in trajectories embedded on a toroidal surface in phase space.

%
%

In this sense, the jitter signature is now more deterministic, but, in the real case, it is 'randomised' due to continuous forcing by stochastic background turbulence. In fact, because the flow is a convective amplifier, the wave-wave interactions can never fully take over and the asymptotic behaviour can never happen because information is continuously convected out of the system. The dynamical systems reproduced by \emph{Model T} and \emph{Model LC} evolve according to initial conditions and a dynamic law that cannot evacuate the information convectively. 

One might imagine the phase space as a dense ensemble of trajectories evolving on limit cycles and toroidal surfaces, associated with wave-wave interactions. Due to the sensitivity of the system to external noise, the system will continuously and erratically deviate from these attracting solutions without setting on any one of them as observed in the spatiotemporal plots and the spectrogram in Sec.~\ref{sec:tseries}. The high sensitivity to external disturbances is inherently connected to the convective nature of the flow. This picture can be linked to recent ideas regarding wavepacket theory: the organisation observed is understood as being due to the resolvent operator forced by background turbulence. In this sense, the resolvent operator is an {\em{organiser}} of the flow, while the turbulence that forces wavepackets via this operator is the {\em{disorganiser}}. 

Also, note that the resolvent analysis has proved to be particularly efficient for $St>0.3$, where the linear wavepackets are in remarkable agreement with the coherent structures observable in the jet flow. On the other hand, at low $St$ -- and particularly at low Mach number -- non-linear wave-wave interactions may be more relevant as the linear analogy fails: this makes our conjecture particularly appealing for explaining this discrepancy.

%
%
\section{Input-output modelling for control: is linear modelling enough?}\label{sec:rom2}
\begin{figure}[t]
\centering 
\subfigure[Input $u(t)$ at $x/D=2.5$]{\includegraphics[width=0.45\textwidth]{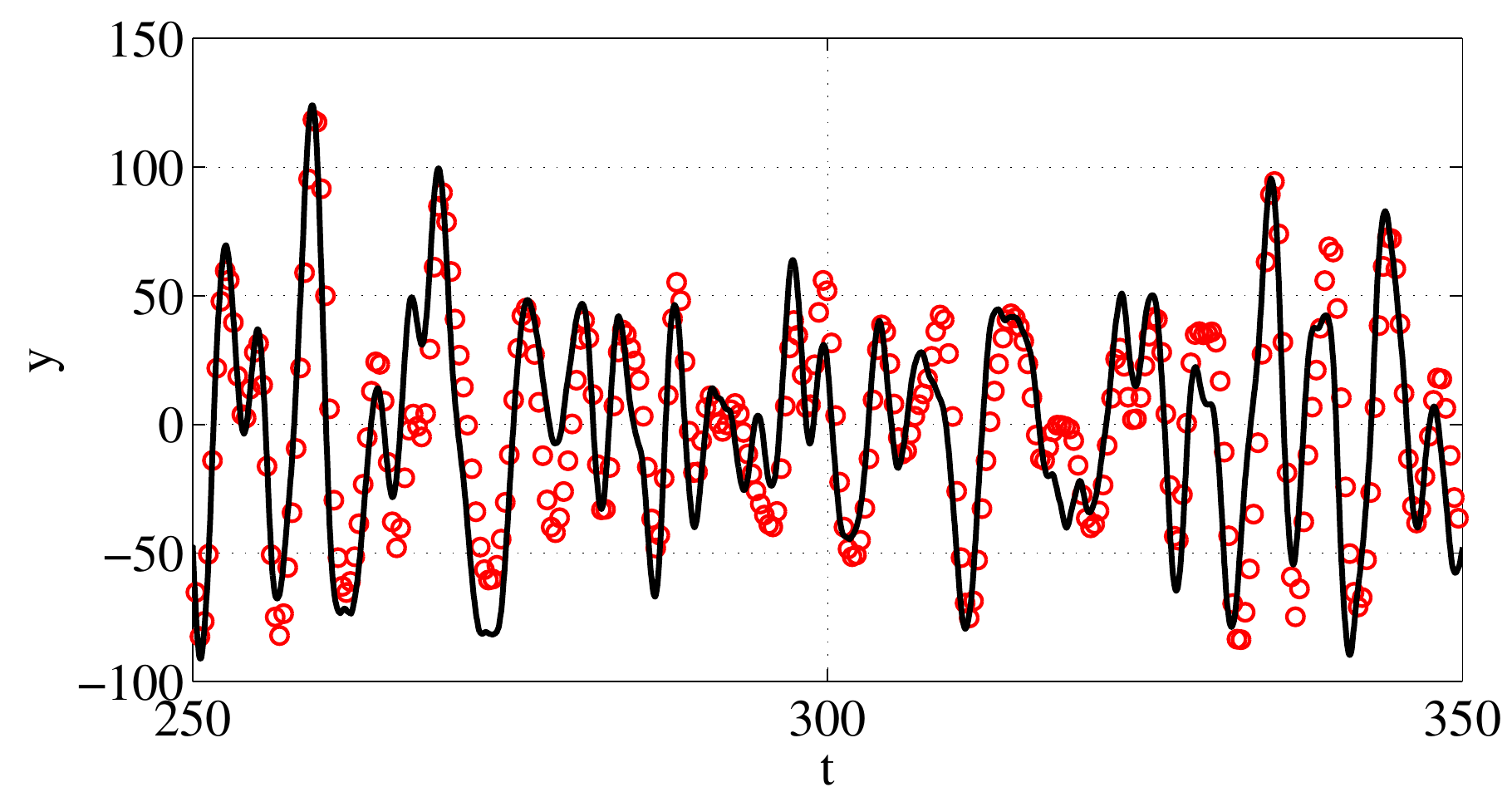}}
\subfigure[Input $u(t)$ at $x/D=3.3$]{\includegraphics[width=0.45\textwidth]{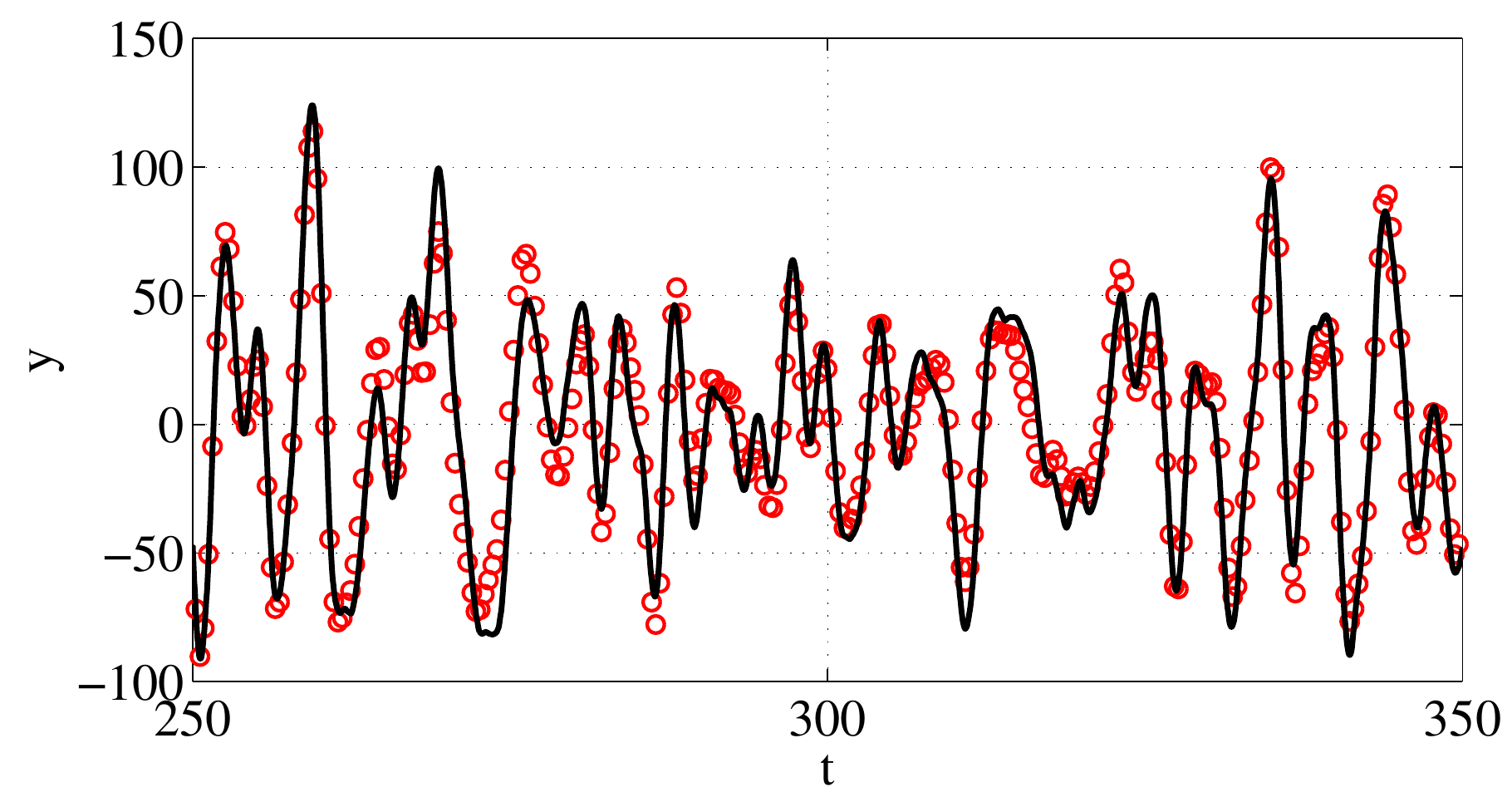}}
\subfigure[Input $u(t)$ at $x/D=4.1$]{\includegraphics[width=0.45\textwidth]{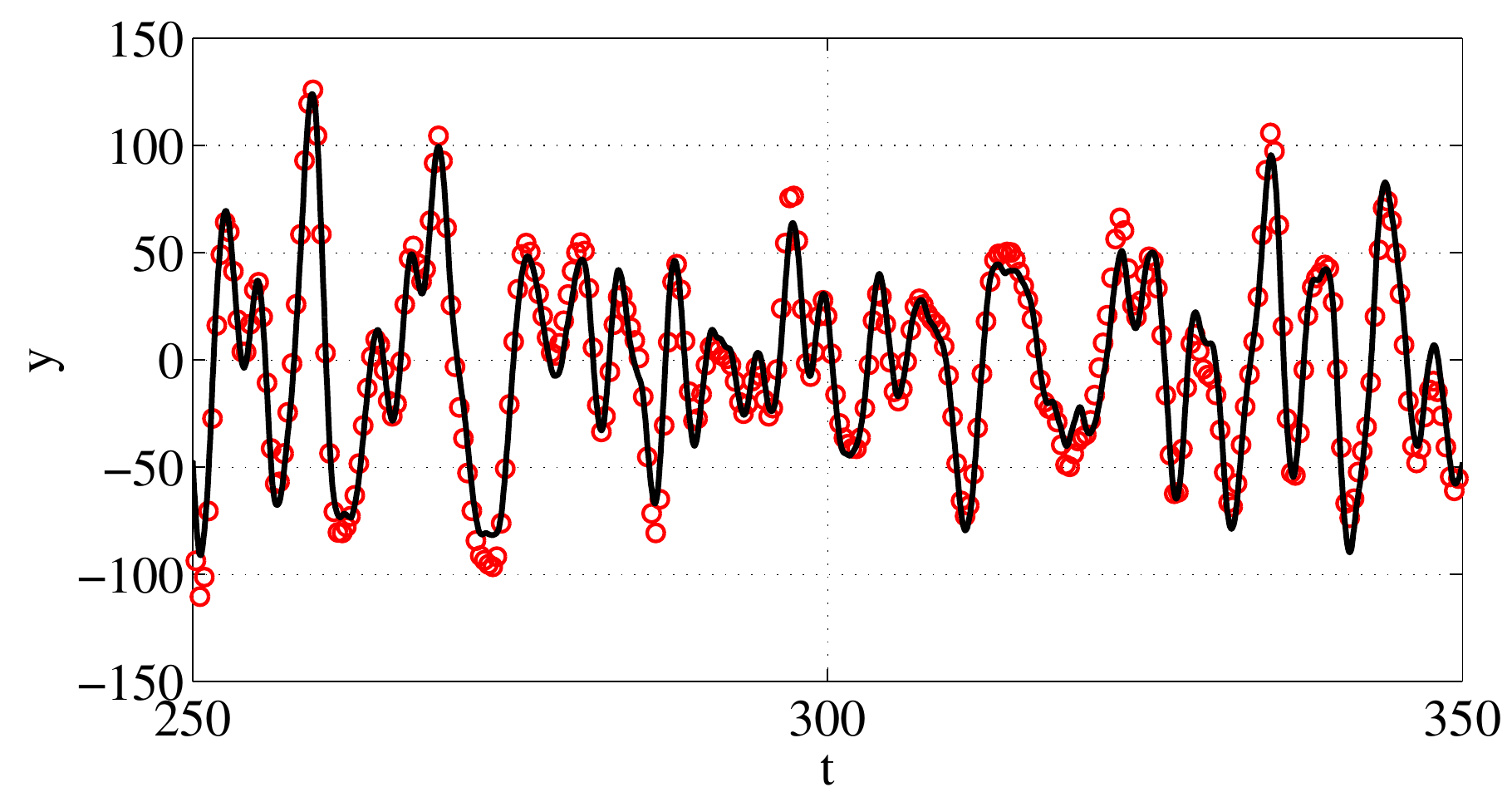}}
\subfigure[Input $u(t)$ at $x/D=4.9$]{\includegraphics[width=0.45\textwidth]{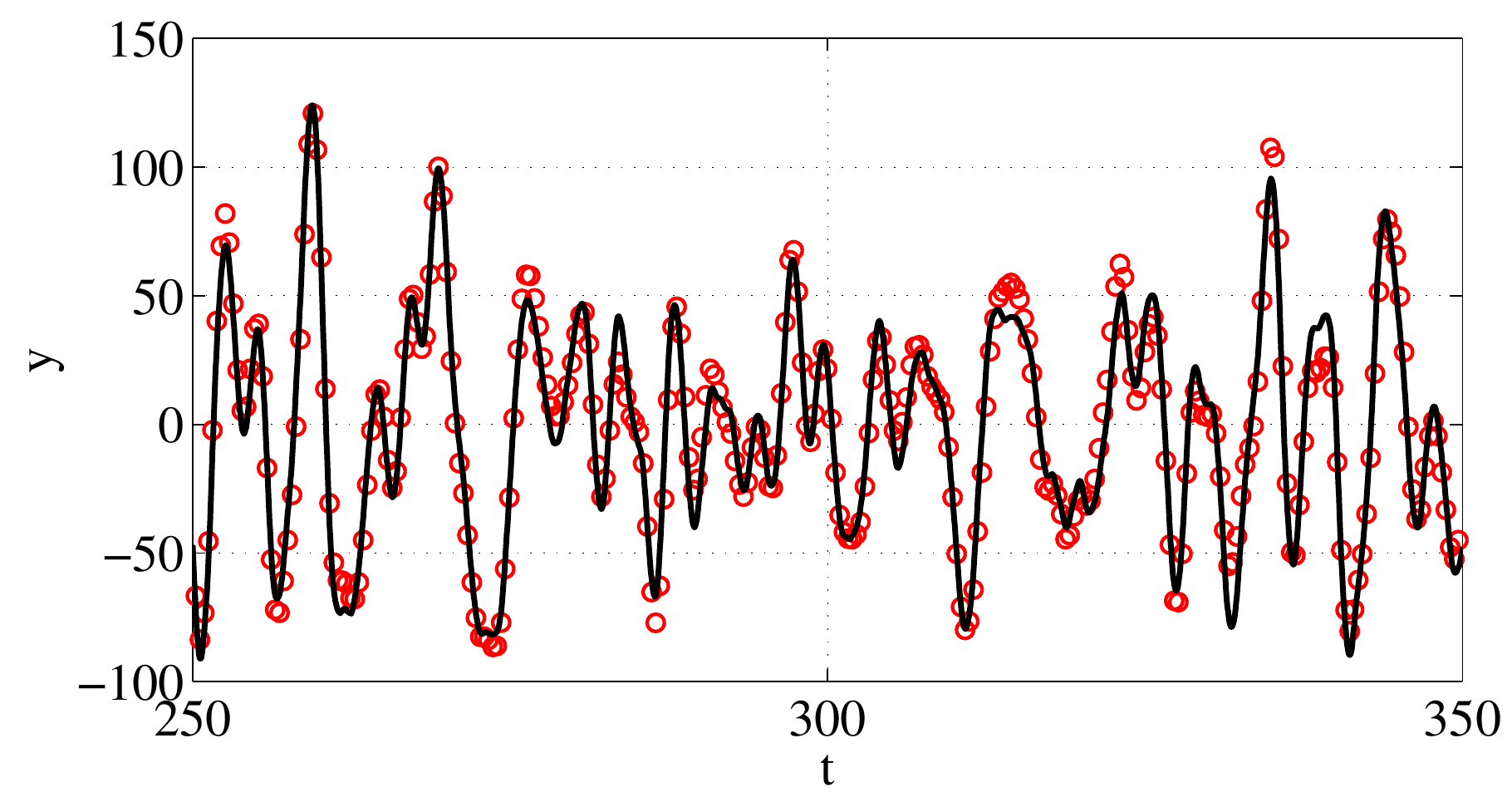}}
\caption{Reduced-order model vs. original time-series. Solid, black lines indicate the original signal at $x/D=5.7$; the red dots indicate the model prediction. Four inputs are considered, at  $x/D=[2.5,\,3.3,\,4.1,\,4.9]$.}
\label{fig:lin_arma1}
\end{figure}

When the forcing term $u(t)$ is represented by an output of the system, used as an input of the model, system identification identifies a transfer function between inputs and outputs. In particular, we use the output at  $x/D=5.7$ and inputs placed in four positions, $x/D=[2.5,\,3.3,\,4.1,\,4.9]$. The models include only one input (single-input-single-output modelling, SISO); the performance are shown in Fig.~\ref{fig:lin_arma1}$a$-$d$, where we compare the model prediction with the original time-series. The window of assimilation is $t\in [0, 164]$. The validation is performed by integrating the model in the window $t\in [0, 620]$.

The application of the full non-linear ARMAX algorithm results in a polynomial expression: the final models are fully linear and based on a limited number of coefficients $\boldmath{\alpha}$ and $\boldmath{\beta}$, $N_{\alpha,\beta}=20$. The good results can be explained by observing the spectral analysis in Fig.~\ref{fig:pwelchseries}. As already noted, the convective nature of the flow implies a progressive drift of the main frequency of amplification and a slight variation of the maximum amplitude in the spectrum as we move further downstream; in this sense, we need only a limited number of degrees of freedom for modulating amplitudes and frequencies, and correctly representing the time-delays of the system. This observation justifies the small dimensions of the models, that here act as filters.

\medskip
The standard deviation between the resulting model and the original time-series is generally low. In Fig.~\ref{fig:lin_arma2}$a$, the results are summarised for a number of outputs along the streamwise direction. In particular, six different positions along $x/D$ are introduced as input, in the range $2.9 \le x/D \le 4.9$, while as outputs are considered the locations downstream of the inputs between $x/D=3.3$ and $x/D=5.7$. Except the transfer functions with inputs at $x/D=2.9-3.3$ and output at $x/D=5.7$, characterised by a standard deviation $\sigma\approx 0.3-0.5$, in all the other cases the standard deviation is below $\sigma=0.3$.  As already mentioned above, all the shown models are linear: this observation has interesting implications when considering the design of linear controllers.

\medskip
These results confirm and extend the findings of \cite{sasaki2015real}, where linear ARMAX modelling, parabolised stability equation (PSE) and empirical transfer function were compared for the estimation of downstream propagating wavepackets, showing equivalent results. With respect to the work of \cite{sasaki2015real}, in our analysis the linearity of the model is a result of the optimisation process and not a working hypothesis: despite the fact that candidate models introduced in the identification procedure were characterised by all the terms of Eq.~\ref{eq:fullnarmax} up to order $n=3$, non-linearities appear to be un-necessary for an optimal representation of the transfer functions.

\subsection{Robustness and control}
\begin{figure}
\centering 
\subfigure[Models' performance]{\includegraphics[width=0.475\textwidth]{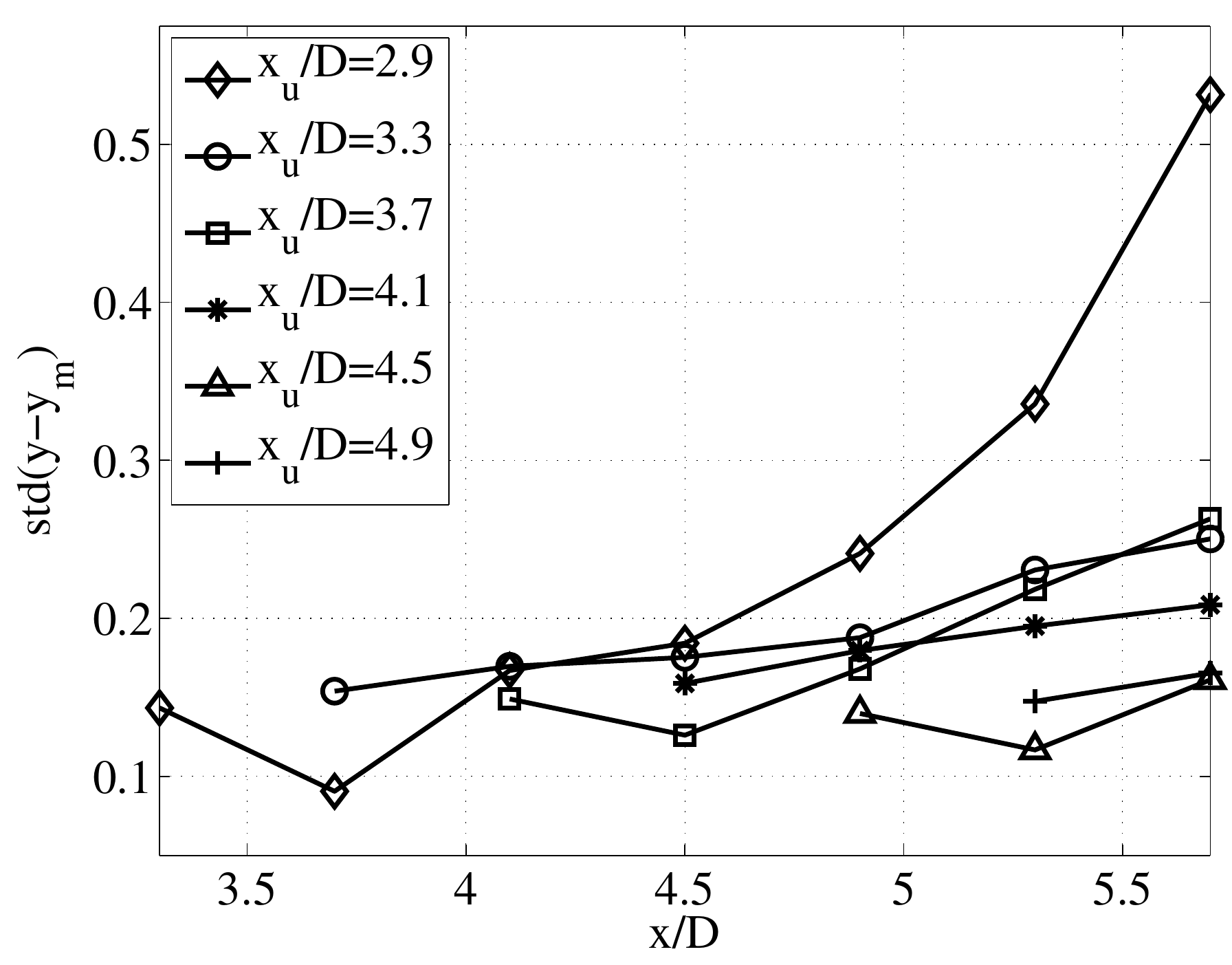}}
\subfigure[Robustness]{\includegraphics[width=0.475\textwidth]{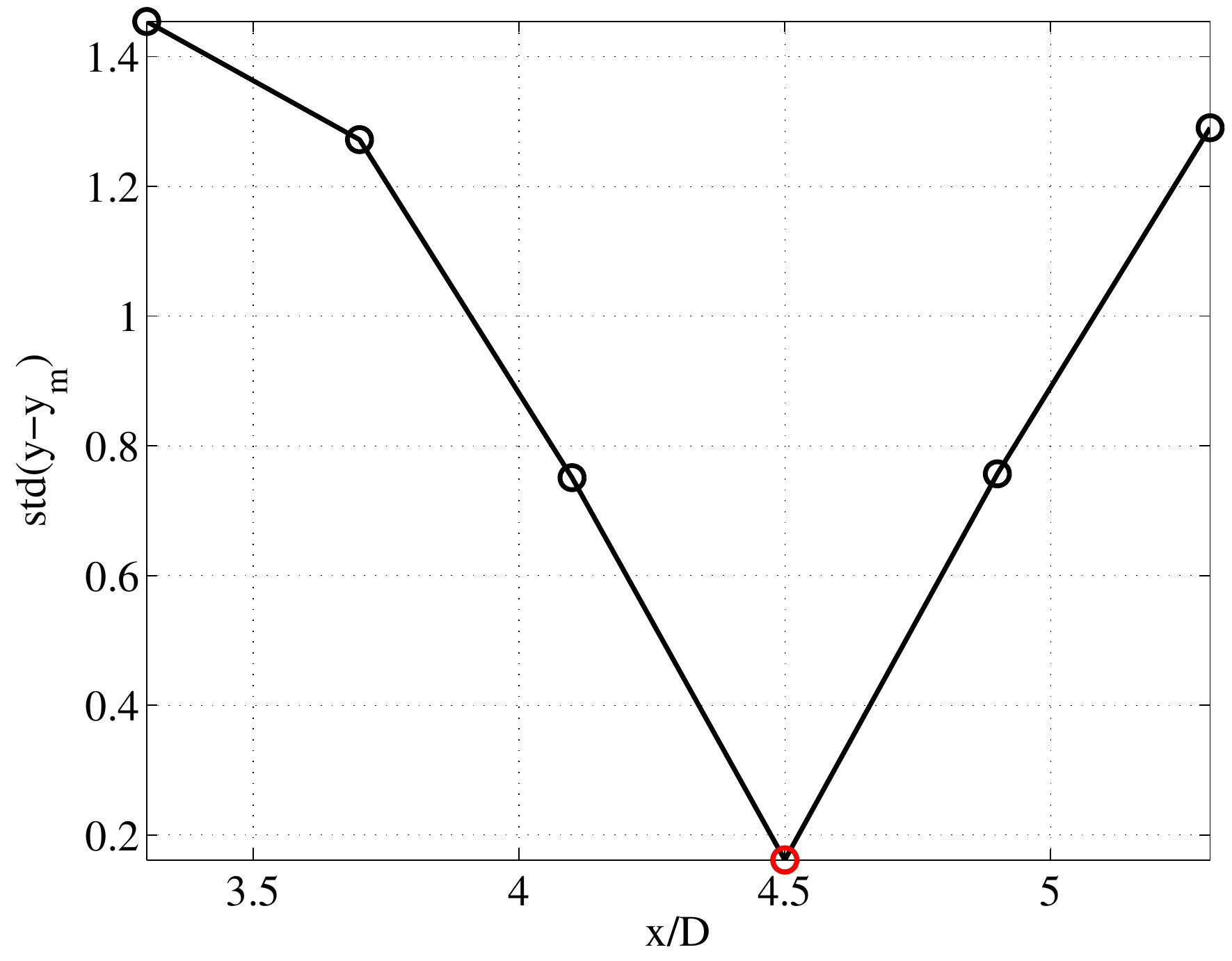}}
\caption{Left: the normalized standard deviation between the models and the time-series is analysed. The $x$-axis indicates the location of the output, while in the legend the six lines correspond to the inputs. In total, $27$ models are analysed, most of which are characterised by a standard deviation $\sigma<0.25$. Right: the model is computed between the input at $x/D=4.5$ and the output at $x/D=5.7$; the same model, is tested by using different inputs between $x/D=3.3$ and $x/D=5.7$; the test shows the lack of robustness with respect to the time-delays of the system.}
\label{fig:lin_arma2}
\end{figure}

The limited amount of dynamical information retained in these models lead to robustness issues. The set of coefficients is obtained as least-squares solution from the assimilation datasets at given amplitude. The transfer function only retains information regarding the time-delays of the system, the modulation of the amplitude and the frequency bandwidth. Any sources of uncertainty can lead to robustness issues. As an example, in Fig.~\ref{fig:lin_arma2}$b$, we illustrate the lack of robustness with respect to time-delays, due to the convective time of propagation of the perturbations. The model assimilated using the input at $x/D=4.5$ is applied with different inputs, taken in the range $3.3\le x/D \le 5.3$. The validity of the models quickly decays. Values of the standard deviation $\sigma>1$ indicate phase-opposition of the resulting model with respect to the original time-series.

The robustness issues can be attacked using different strategies. Control design can be conceived by accounting for the robustness issues arising from uncertainties of the system within the robust framework or the adaptive control theory (see \cite{sipp2016linear,fabbiane2014adaptive}). In particular, with respect to the convective nature of the flow, feed-forward schemes combined with self-tuning, adaptive controllers can be applied. These controllers are characterised by two time-scales: a \emph{fast} time-scale, for the real-time estimation of the input $u(t)$, and a \emph{slow} time-scale, based on feedback measurements of the flow necessary for correcting off-design conditions. Examples of these strategies can be found for weakly non-linear transitional flows (see \cite{semeraro2013transition}) or non-linear, turbulent cases (see \cite{rathnasingham2003active}). 

\section{Conclusions}\label{sec:conclu1}
Coherent large-scale structures in turbulent jets -- usually referred as wavepackets -- have been subject of numerous investigations for the role that such structures might play in sound radiation. Remarkably, it is well established that certain aspects of the wavepacket dynamics can be modelled by means of linear $ansatz$ (\cite{jordan2013wave}). In this work, we considered an unforced, isothermal, subsonic jet issuing from a round nozzle with a fully turbulent boundary layers at $Re - 5.7\times 10^5$ and $Ma=0.6$. The temporal behaviour of the axisymmetric wavepackets is analysed; starting from the studies by \cite{breakey2013near}, we investigate the experimental, spatio-temporal measurements from a dynamical point of view.  To the best of our knowledge, this is the first time such an analysis has been attempted. The key point is the remarkable spatio-temporal coherence of the measurements: despite the high Reynolds number, the low-dimensional projection of the dynamics reveals a well-organised behaviour. This aspect raises questions regarding the temporal behaviour of the wavepackets, the nature of the non-linearities involved and the possibility of designing control strategies starting from this knowledge.

\medskip
In order to answer these questions, a variety of strategies ranging from statistical tools to system identification have been adopted. 
The estimation of the correlation dimension and the embedding dimension of the dynamical system confirmed that the organised temporal behaviour is low-dimensional and deterministic. The minimal embedding ranges between $m=10$ and $m=15$, while the correlation dimension is rather low at locations corresponding to the end of the potential core, $d_2\approx 6$. These dimensions are nonetheless still too large for useful characterisation of attractor geometry.

\medskip
 For this reason, alternative strategies are necessary for {exploring} the dynamics of the system. A system identification approach is pursued. The time-series taken at a streamwise location of the jet is assumed as output; we replace the original dynamics with surrogate, autonomous models using data-assimilation based on this output. These models are limited, as they reproduce only a part of the dynamics and are influenced by the choice of the parameters.

By comparing the results from three different classes of models, we suggest that the jittering dynamics might be described as a combination of extrinsic and intrinsic mechanisms at work. We observe that our surrogate models reproduce the ideal case where the background turbulence is not active; in this limit, the dynamics of the system would be dominated by wave-wave interactions leading to stable limit cycles and/or solutions lying on toroidal attractors. Evidence on this behaviour is provided by the spectral analysis of the transient dynamics of these solutions that qualitatively reproduces the local (in time) dynamics of the original system. However, in the real case, the extrinsic dynamics drives the system: wave-wave interactions can never fully take over and the asymptotic behaviour of the ideal models can never happen. This interpretation justifies the self-organised features appearing in the time-frequency analysis of the time-series and it is consistent with the nature of the jets as noise-amplifiers: due to the sensitivity of the system to external noise, the trajectory in phase space will continuously escape from these attracting solutions.

\medskip
Given these interpretations, control strategies can be conceived. Due to the convective nature of the system and its implications regarding the non-linear interactions, system identification is used for computing non-autonomous models, where the inputs are represented by local measurements taken upstream of the outputs. These models are filters that i) account for the time-delays and ii) modulate the amplitudes and the frequencies already existing in the inputs. In particular, we found that non-linearities are un-necessary for an optimal representation of these models. In other words, if the inputs contain a large part of the dynamics, a linear filter is enough to reproduce the temporal behaviour at downstream locations due to the convective nature of the system. The main limit is the robustness of these linear filters; however, in our opinion, this might be more a problem of controllability of the flow, rather than a lack observability and predictability. Thus, we believe that the robustness limitations may be tackled in the control design, by applying strategies such as adaptive control or loop-shaping robust control, and verifying the effects of the actuation on the real flow.

\medskip
Future work should be devoted to further improving the non-linear models for the dynamical analysis. While accepting the impossibility of capturing all the features of the system in low-dimensional models, a possible path to improve our understanding of the jittering would be the identification of surrogate models where a stochastic excitation is also included; in principle, this strategy would help in separating the instrinsic mechanisms from the extrinsic ones. In this sense, we believe that machine learning and statistical learning techniques may be robust alternatives to classical non-linear system identification. Indeed, these techniques are quickly spreading in the community as a powerful tool for system identification (see \cite{SINDy}). 

\bigskip
The authors wish to acknowledge Luis~A.~Aguirre for sharing his system identification package and Christophe~Letellier for fruitful discussions. This work is supported by the Agence Nationale de la Recherche (ANR) under the "Cool Jazz" project, grant number ANR-12-BS09-0024.

\newpage

\appendix

\section{System identification using polynomials expansion}\label{sec:appA}
We introduced in Sec.~\ref{sec:sysID} the basic ideas behind the application of system identification for the qualitative analysis of dynamical system and the definition of control-oriented models. The aim of the section is to provide a quick overview of the algorithms.

We rely on an input-output description of time-invariant systems. The aim of our models is reproducing the behaviour of the output $y(t)$ by defining a non-linear model of the dynamics of the measurement or a transfer function between the input $u(t)$ and the output $y(t)$. For sake of conciseness, we consider a single-input-single-output (SISO) case; nonetheless, the extension to the multi-inputs-multi-outputs cases is rather straightforward from the theoretical point of view.

By assuming a SISO system, we can generalise the equation error models in terms of Volterra series
%
%
\begin{eqnarray}\label{eq:volterra_app}
y(t) = y_0
&+&\sum_{i=1}^{n_\alpha}\alpha_{(1,i)} y(t-i) 
+\sum_{i=1}^{n_\beta}\beta_{(1,i)} u(t-i) 
+\sum_{i=1}^{n_\alpha}\sum_{j=1}^{n_{\beta}}\eta_{(1,i,j)}y(t-i)u(t-j) \nonumber \\
&+&\sum_{i=1}^{n_\alpha}\sum_{i=j}^{n_\alpha}\alpha_{(2,i,j)} y(t-i) y(t-j) 
+\sum_{i=1}^{n_\beta}\sum_{i=j}^{n_\beta}\beta_{(2,i,j)} u(t-i) u(t-j) \nonumber \\
&+& e(t) +... + O(2)
\end{eqnarray}
where $\alpha_i,\,\beta_i,\,\eta_i$ are the $i-th$ order Volterra kernel. The error can be modelled as a moving-average of (unknown) white-noise $w(t)$
\begin{equation}
e(t) = w(t) + \sum_{i=1}^{n_\gamma}\gamma_i w(t-i)
\end{equation}
where $\gamma_i$ are the coefficients for the error modelling. The kernel's coefficients and the error $e(t)$ are the unknowns of our problem. The Volterra-series are an extension of the linear convolution integral and represent non-linear systems as a series of multi-summations (or integrals) of the Volterra kernels and the inputs \cite{billings2013nonlinear}. We can note that the complete model consists of a linear combination of three groups of terms
\begin{enumerate}
\item the auto-regressive terms related to the output $y(t)$;
\item the exogenous terms due to the input $u(t)$;
\item the moving-average description of the error $e(t)$.
\end{enumerate}
Additionally, the non-linear combinations are introduced in the full model. 

\subsection{Solution of the identification problem: a basic ARMAX algorithm}
The first step of the identification procedure consists in a the preliminary computation of the coefficients $\boldsymbol{\theta}^\star=[\alpha_1\,\alpha_2\,... \alpha_n\,\beta_1\,\beta_2\,...\beta_n]$. To this end, we assume that the error is null, such that the least-squares regression can be performed. Note that the error $e(t)$ is modelled as a moving average of white noise $w(t)$, leading to a coloured noise with expected value $\mathbb{E}\left[\boldsymbol{e}(t)\right] \neq 0$; the estimate of the coefficients $\boldsymbol{\theta}$ will be affected by a bias, due to the expected value of the error $\boldsymbol{e}(t)$. In particular, the error is computed as
\begin{eqnarray}
\boldsymbol{e}(t) = \boldsymbol{y} -H_N \boldsymbol{\theta}^\star,
\end{eqnarray}
where $H_N$ is a rectangular Hankel matrix including inputs $\boldsymbol{u}$ and outputs $\boldsymbol{y}$. The Hankel matrix has the dimensions of the number of the coefficients and the window of assimilation. We can introduce a matrix $Z_N$ of the same dimensions as the Hankel matrix $Z_N$ such that
\begin{eqnarray}
\mathbb{E}\left[Z_N \boldsymbol{y} -Z_NH_N \boldsymbol{\theta}^\star\right]=\boldsymbol{0}.
\end{eqnarray}
Thus, with a proper choice $Z_N$, the ``de-biased'' estimate of $\boldsymbol{\theta}^\star$ can be computed as
\begin{eqnarray}
\boldsymbol{\theta}^\star = \left(Z_N^TH_N\right)^{-1}Z^T \boldsymbol{y},
\end{eqnarray}
if $Z_NH_N$ is invertible. This generalisation of the least-squares problem is known as \emph{Instrumental Variable (IV)} method and allows a first estimation of the coefficients in the expansion, while neglecting the error. 

The IV method is iterative. For the first iteration, the matrix of instruments is written as $Z_N=\left[H_N(\eta) \, {H^{(u)}}_n(u)\right]$; a rather effective choice consists in choosing $\boldsymbol{\eta}(t) = \boldsymbol{u}(t-n_d)$, \emph{i.e.} the delayed input. These values are relaxed till convergence.

\medskip
The IV method allows to compute the coefficients $\boldsymbol{\alpha}$ and $\boldsymbol{\beta}$, reducing the effects of the bias due to the non-whiteness of the error $e(t)$. The unknowns of the moving-average part, namely the coefficients $\boldsymbol{\gamma}$ and the white-noise series $w(t)$ can be extracted using appropriate algorithms (see \cite{guidorzi2003multivariable} or \cite{semeraro2015techrep}).

\medskip
Thus, we can divide the basic algorithm into two steps:
\begin{enumerate}
\item Evaluate the coefficients $\boldsymbol{\alpha}$ and  $\boldsymbol{\beta}$ using the iterative IV method.
\item Apply the moving-average algorithm for defining the coefficients $\boldsymbol{\gamma}$ and the white-noise $w(t)$.
\end{enumerate}
This prototypical algorithm allows a first estimate of the kernel coefficients and the error of the moving average description. We can improve these estimates using the \emph{extended least-squares} (ELS) approach; in particular, we can consider the first estimation of the error as an input of the system: in this way, the estimates will be de-biased by explicitly accounting the error. Moreover, an iteration can be cast: the difference between the prediction and the original output, allows to refine again the kernels estimate and further improve the $e(t)$ prediction.

\subsection{Other techniques}
The main difference between the linear modelling and the full non-linear Volterra series is mainly related to the dimensions of the problem and the risk of ill-conditioness. Already cubic non-linearities may introduce nearly null entries, leading to rank-deficiency and singularities. To this end, the regression step can be replaced by well-conditioned algorithms, such as the \emph{orthogonal least-squares} (OLS); this iterative scheme introduces in the least-squares process a Gram-Schmidt orthogonalisation, which ensures that each new column added in the matrices is orthogonal to all previous columns. Moreover, this process allows us to estimate the relative relevance of the kernel's coefficient in residual terms; thus, elements that are not relevant are discarded during the process. The OLS approach is described by \cite{billings2013nonlinear} or \cite{aguirre2009modeling}, and allows the identification of minimal non-linear models by keeping only the relevant terms of the model. This is the strategy implemented in the routines adopted in our manuscript. 

 Other methods for the system identification are based on the minimisation of the prediction at a given time-horizon, based on gradient-based optimisation process; an example is given by the \emph{predictive error method} (PEM), implemented in the \emph{system identification toolbox} by \cite{ljung2007system} (see also \cite{ljung1999system}).

\newpage

%
%
\section*{References}
\bibliographystyle{elsarticle-num}
\bibliography{biblio_qualdyn}

\end{document}